\documentclass[12pt]{article}

\usepackage[margin=1in]{geometry} 
\usepackage{amsmath,amsthm,amssymb}
\usepackage{graphicx}
\usepackage{tensor}
\usepackage{cancel}
\usepackage{multicol}
\usepackage{graphicx}  % Required for including images
\usepackage{caption}   % For better caption formatting
\usepackage{soul}
\usepackage{slashed}
\usepackage{subcaption}
\usepackage{hyperref}
\usepackage{subcaption}
\usepackage{comment}

\begin{document}

\begin{titlepage}

\begin{flushright}
{}
%arXiv Number (to be placed)
\end{flushright}
\vskip 2.5cm

\begin{center}
{\Large \bf Primordial Gravitational Wave Birefringence in a\\
De Sitter Background with Chern-Simons Coupling}
\end{center}

\vspace{1ex}

\begin{center}
{\large Abhishek Rout\footnote{{\tt arout@email.sc.edu}}
and Brett Altschul\footnote{{\tt altschul@mailbox.sc.edu}}}

\vspace{5mm}
{\sl Department of Physics and Astronomy} \\
{\sl University of South Carolina} \\
{\sl Columbia, SC 29208} \\
\end{center}

\vspace{2.5ex}

\medskip

\centerline {\bf Abstract}

\bigskip

In this work, we investigate tensor perturbations in a de Sitter background within the framework of Chern-Simons
modified gravity. We introduce transverse-traceless perturbations and analyze how the Chern-Simons Cotton tensor
induces parity-violating modifications to gravitational wave propagation, while the Pontryagin density vanishes
at linear order. Using a mode decomposition of the scalar background field, we derive the sub- and super-horizon
limits of the wave equations and uncover chiral corrections in the dispersion relations of tensor modes. The
resulting birefringence exhibits both amplitude and velocity components, alternating with the phase of the scalar
field. Particular solutions sourced by the scalar background show helicity-dependent amplification and a
characteristic scaling of the radiated flux that reduces smoothly to the Minkowski limit. The accumulated phase
difference between right- and left-handed modes grows quadratically inside the horizon and becomes frozen
outside, leaving a permanent parity-violating imprint in the primordial tensor spectrum. Finally, by promoting
the Chern-Simons field to a massive dark matter candidate, we demonstrate how its mass-dependent dynamics connect
gravitational birefringence to axion-like dark matter phenomenology.

\bigskip

\end{titlepage}

\section{Introduction}
\label{sec-intro}

The study of perturbations in de Sitter spacetime occupies a central role in modern 
theoretical physics and cosmology. The de Sitter metric, first introduced as an exact 
solution of Einstein’s equations with a positive cosmological constant \cite{ref-desitter}, 
remains one of the simplest yet most profound models of a universe dominated by vacuum energy. 
Historically, it provided one of the earliest connections between general relativity and 
cosmology, and later it became a canonical background for understanding quantum fields in 
curved spacetimes and the large-scale dynamics of the universe~\cite{ref-hawking,ref-mukhanov}.

From the perspective of studies of the early universe, inflationary cosmology posits a period of 
exponential expansion driven by a scalar field vacuum energy, approximating a quasi-de Sitter 
background to the extent that the energy in question remains relatively constant.
This idea was pioneered in the early 1980s~\cite{ref-guth,ref-linde,ref-albrecht}, 
with subsequent work showing how quantum fluctuations that were stretched out during inflation could seed 
the inhomogeneities that lead to observable anisotropy in
the cosmic microwave background (CMB) and the large-scale structure of the 
universe~\cite{ref-mukhanov_a,ref-starobinsky,ref-bardeen}. The analysis of linear perturbations 
in de Sitter space therefore forms the backbone of modern cosmology, underlying predictions for 
CMB polarization, primordial gravitational waves, and the formation of cosmic structures. 
These predictions are now tested with exquisite precision by observational missions such as 
Planck and BICEP/Keck~\cite{ref-planck,ref-bicep}.

At late times, the observed accelerated expansion of the
universe~\cite{ref-reiss,ref-perlmutter} can also be effectively modeled by a nearly de Sitter geometry 
with a small positive cosmological constant, interpreted as dark energy. To the extent that the dark energy
density remains constant over time (like a pure cosmological constant), the universe will become
more like de Sitter space over time, because all other sources of mass energy become more rarefied
in the course of the Hubble expansion. This dual relevance---to
both the very early and the very late universe---makes perturbations of de Sitter space 
uniquely important for bridging between fundamental theory and precision cosmological data.

Beyond general relativity, many modified gravity theories have been proposed to address 
longstanding conceptual issues such as the cosmological constant problem, the unification of gravitation with 
quantum mechanics, and the possibility of parity violation in fundamental interactions. Among 
these, Chern-Simons (CS) modified gravity has attracted significant attention. Originally 
emerging in the mathematical study of topological invariants in $2+1$ dimensions~\cite{ref-chern}, and later
applied to $(3+1)$-dimensional
gravity by Jackiw and Pi~\cite{ref-jackiw}, CS gravity extends the Einstein-Hilbert action 
by coupling a scalar field to the Pontryagin density, thereby introducing parity-violating 
corrections. This framework has since been extensively developed in astrophysics and cosmology, 
from black hole perturbations ~\cite{ref-yunes2009,ref-yunes2010} to gravitational wave
propagation~\cite{ref-alexander}, to possible cosmological imprints~\cite{ref-lue,ref-contaldi}.

Studying perturbations of de Sitter space in the CS framework provides a particularly clean 
arena for probing how parity-violating terms affect cosmological stability and the dynamics of 
primordial fluctuations. Of special importance is the Pontryagin constraint, which arises from 
the variation of the CS action. Recent analyses have clarified that this constraint vanishes 
at first order in linearized gravitational wave perturbations on de Sitter backgrounds~\cite{ref-dyda},
ensuring the consistency of de Sitter space as a cosmological solution in this 
framework. This suggests that parity-violating corrections are likely to appear at higher orders 
or in less symmetric spacetimes.

The motivation for introducing a CS term in a cosmological setting is to
capture possible parity-violating effects in gravity that are expected in several high-energy
extensions of general relativity, including string-inspired and axion-like frameworks. In the
early universe, where curvature scales are large and scalar fields are naturally dynamical,
such terms can leave imprints on primordial gravitational waves without spoiling the background
de Sitter dynamics. The resulting helicity-dependent propagation provides a clean
and potentially observable signature of new gravitational physics, linking fundamental parity
violation to cosmological observables such as the primordial tensor spectrum and CMB
polarization.

In this paper, we explicitly study linearized perturbations of the de Sitter metric within
CS-modified gravity. By deriving and analyzing the perturbed field equations, we evaluate
the Pontryagin constraint and show that it vanishes to first order. These results clarify the precise role
of possible CS terms in cosmological models, provide insight into how parity-violating effects could
manifest during inflation, and point toward possible observational signatures in future gravitational
wave and CMB experiments. The paper is organized as follows. In sec.~\ref{sec:CS}, we review the
CS-modified action and the resulting field equations. In sec.~\ref{sec:Perturb}, we introduce
tensor perturbations of the de Sitter metric and obtain the corresponding perturbed field equations. The
Pontryagin constraint is evaluated in sec.~\ref{sec:Pontryagin}, where we demonstrate its vanishing at
linear order. In sec.~\ref{sec:Scalar}, we solve for the scalar background field in de Sitter spacetime
and discuss its sub- and super-horizon behavior. Sec.~\ref{sec:Solutions} presents the solutions to
the perturbed field equations, including homogeneous and particular modes, along with the resulting
birefringence effects. The associated energy flux and its reduction to the flat-spacetime limit are
analyzed in sec.~\ref{sec:Flux}. In sec.~\ref{sec:Phase}, we calculate the accumulated phase difference
between helicity states, highlighting its distinct scaling in the two horizon regimes. Finally,
in sec.~\ref{sec:DM} we extend the framework by promoting the CS scalar to a massive
dark matter candidate, before summarizing our results and their observational implications in
sec.~\ref{sec:Conclusion}.

\section{CS-Modified Gravity}
\label{sec:CS}

The CS-modified action is given by
\begin{equation}
    \mathcal{S} = \frac{1}{16\pi G}\int d^4x \sqrt{-g} \left[\mathcal{R} 
    + \frac{\vartheta}{4}(^{\star}\mathcal{R}\mathcal{R}) - \frac{1}{2}(\nabla\vartheta)^2 
    - V(\vartheta) + \mathcal{L}_{\text{mat}}\right],
    \label{eq-actionS}
\end{equation}
where the quantity $^\star \mathcal{R}\mathcal{R}$ is known as the Pontryagin density and is defined to be
\begin{equation}
    ^{\star}\mathcal{R}\mathcal{R}
    = \tensor{^{\star}R}{^{\sigma}_{\tau}^{\mu\nu}}\tensor{R}{^{\tau}_{\sigma\mu\nu}};
\end{equation}
the dual of the Riemann tensor is
\begin{equation}
    \tensor{^{\star}R}{^{\sigma}_{\tau}^{\mu\nu}}
    = \frac{1}{2}\varepsilon^{\mu\nu\alpha\beta}\tensor{R}{^{\sigma}_{\tau\alpha\beta}}.
\end{equation}
We note that an explicit cosmological constant term has been omitted for brevity. Throughout
this work, its effect is implicitly incorporated via the scalar potential, which supports
the de Sitter background considered below in sec~\ref{sec:Perturb}.

Varying this action with respect to the CS scalar field $\vartheta$ gives us the wave equation for the
scalar,
\begin{equation}
\label{eq-theta-motion}
    \Box\vartheta - V'(\vartheta) = -\frac{1}{4}(^{\star}\mathcal{R}\mathcal{R}),
\end{equation}
where $\Box$ is the standard massless D'Alembertian, defined in curved spacetime to be
\begin{equation}
    \Box = \frac{1}{\sqrt{-g}} \partial_\mu \left(\sqrt{-g}\, g^{\mu\nu} \partial_\nu\right).
\end{equation}
Meanwhile, varying the action with respect to the metric tensor $g_{\mu\nu}$ gives us the modified Einstein
field equations,
\begin{equation}
    G_{\mu\nu} + C_{\mu\nu} + \Lambda g_{\mu\nu}
    = 8\pi G\left[T^{(\vartheta)}_{\mu\nu} + T^{(\text{mat})}_{\mu\nu}\right],
\end{equation}
where we have introduced the Cotton tensor $C_{\mu\nu}$, defined as
\begin{equation}
    C^{\mu\nu} = -\frac{1}{2}\left[(\nabla_{\alpha}\vartheta)\,
    \tensor{\epsilon}{^{\alpha\beta\gamma(\mu}}\nabla_{\gamma}\tensor{R}{^{\nu)}_{\beta}} 
    + (\nabla_{\alpha}\nabla_{\beta}\vartheta)\,^{\star}\tensor{R}{^{\beta\, (\mu\nu)\,\alpha}}\right];
\end{equation}
we have used $\epsilon^{\alpha\beta\gamma\mu} = \frac{\varepsilon^{\alpha\beta\gamma\mu}}{\sqrt{-g}}$,
with $\varepsilon^{\alpha\beta\gamma\mu}$ being the Levi-Civita symbol. As usual, parentheses around
indices indicate symmetrization.

The stress-energy tensor for the new field $\vartheta$ that modifies the theory takes the form
\begin{equation}
    T^{(\vartheta)}_{\mu\nu} = (\nabla_{\mu}\vartheta)(\nabla_{\nu}\vartheta)
    - \frac{1}{2}g_{\mu\nu}\left[(\nabla\vartheta)^2 + V(\vartheta)\right],
\end{equation}
typical for a scalar.
In case of vacuum solutions [meaning the vanishing $T^{(\text{mat})}_{\mu\nu} = 0$ of the portion of
the stress-energy tensor that refers to conventional matter], we have the field
$\vartheta$ as the only gravitational source, and the Einstein equations reduce to
$G_{\mu\nu} + C_{\mu\nu} + \Lambda  g_{\mu\nu} = 8\pi G\,T^{(\vartheta)}_{\mu\nu}$.

\section{Perturbed Field Equations}
\label{sec:Perturb}

Our perturbed metric for the de Sitter space is
\begin{equation}
    ds^2 = \alpha(\eta)^2\left[-d\eta^2 + \left(\delta_{ij} + h^{TT}_{ij}\right)\,dx^i\,dx^j\right],
\end{equation}
where $\alpha(\eta)$ is the Robertson-Walker scale factor, which for the de Sitter spacetime takes
the forms $\alpha(t)=\alpha_{0}e^{Ht}$ and $\alpha(\eta)=l/\eta$ in terms of the coordinate
and conformal times, respectively;
and where $h^{TT}_{ij}$ contains the transverse, traceless (TT) tensor modes corresponding to
gravitational waves. A single pair of TT tensor mode are conventionally written
\begin{equation}
    h^{TT}_{ij} = \begin{bmatrix}
        h_{+} & h_{\times} & 0\\
        h_{\times} & -h_{+} & 0\\
        0 & 0 & 0\\
    \end{bmatrix},
\end{equation}
describing a wave propagating in the $z$-direction [i.e. $h_{+} = h_{+}(\eta - z)$ and
$h_{\times} = h_{\times}(\eta - z)$].
We restrict attention to the transverse, traceless tensor modes, which correspond to the
propagating gravitational wave degrees of freedom and which are free of gauge ambiguities. In
CS gravity these modes would provide the cleanest and most direct probe of parity
violation, since the Cotton tensor induces chirality-dependent corrections already at linear
order. Scalar and vector perturbations, which involve additional gauge subtleties and mode
mixing, are left for future work.

Using this form of the perturbation in the metric and making the
assumption that our scalar field may be chosen to be of the form $\vartheta = \vartheta(\eta, z)$
to mimic the wave perturbations, we get the field equations of the form 
\begin{eqnarray}
        \label{eq-1+}
        \frac{1}{2}\Box h_{+} + \frac{1}{\alpha^2}C\left[h_{\times}\right] & = & S(\eta,z)\\
        \label{eq-1-}
        \frac{1}{2}\Box h_{\times} + \frac{1}{\alpha^2}C\left[h_{+}\right] & = & S(\eta,z),
\end{eqnarray}
where we have
\begin{equation}
    S(\eta,z) = 4\pi G\left[\left(\partial_{\eta}\vartheta\right)^2
    - \left(\partial_{z}\vartheta\right)^2\right],
\end{equation}
which acts as the source term for the waves, and the term $C[h_{\bullet}]$ takes care of the Cotton tensor's
effects. The form of $C[h_{\bullet}]$ is
\begin{equation}
    C[h_{\bullet}] = \left[\frac{v_{\eta}}{2}\partial_{z}\Box - \frac{v_{z}}{2}\partial_{\eta}\Box
    - \frac{4v_{z}}{\eta}\Box + v_{\eta\eta}\left(\frac{1}{2}\partial_{z}\partial_{\eta}\right)
    + v_{\eta z}\left(\frac{1}{2}\Box\right)
    + v_{zz}\left(-\frac{1}{2}\partial_{\eta}\partial^{z}\right)\right]h_{\bullet}.
\label{eq-1-bullet}
\end{equation}
We have used the definitions $v_{\alpha} = \partial_{\alpha}\vartheta$ and
$v_{\alpha\beta} = \partial_{\alpha}\partial_{\beta}\vartheta$ for the first and second derivatives of the
scalar field. The notation $h_{\bullet}$ indicates that it is indeterminate which mode (and indeed, which
mode basis for the propagating tensor perturbation modes) appears.

\subsection{Operator diagonalization and decoupling of tensor modes}
\label{sec:diagonalization}

It is important to treat the Cotton contribution as a linear \emph{operator} acting on the two polarization
components, not as an ordinary scalar coefficient.  To make this explicit we collect the two polarization
equations [eqs.\ \eqref{eq-1+}--\eqref{eq-1-}] into a single operator equation for the two-vector
\[
\mathbf h(\eta,z)\equiv\begin{pmatrix} h_+(\eta,z) \\[4pt] h_\times(\eta,z)\end{pmatrix}.
\]
In operator form the coupled system may be written as
\begin{equation}\label{eq:operator-system}
\frac12\Box\,\mathbf h(\eta,z) + \frac{1}{\alpha^{2}(\eta)}\,\mathcal C\,\mathbf h(\eta,z) \;=\; \mathbf S(\eta,z),
\end{equation}
where \(\mathbf S\) denotes the corresponding two-component source and $\mathcal{C}$ is the $2\times2$
operator matrix whose off-diagonal structure follows from the component form of the Cotton term [cf.\ eq.\
\eqref{eq-1-bullet}].  From that component structure one obtains
\begin{equation}\label{eq:C-matrix}
\mathcal C \;=\; \begin{pmatrix} 0 & C \\ C & 0 \end{pmatrix},
\end{equation}
with \(C\) the linear Cotton operator which contains derivatives and factors of the background scalar
field $\vartheta$.  (Crucially, $C$ is an operator, not a commuting scalar.)

To diagonalize $\mathcal{C}$, we perform the orthogonal basis change
\begin{equation}\label{eq:T-def}
T \;=\; \frac{1}{\sqrt2}\begin{pmatrix}1 & 1\\[4pt] 1 & -1\end{pmatrix},\qquad
\mathbf h_{\mathrm s} \equiv T\,\mathbf h = \begin{pmatrix} h_+^{(\mathrm s)} \\[4pt]
 h_-^{(\mathrm s)}\end{pmatrix}
\;=\; \frac{1}{\sqrt2}\begin{pmatrix} h_+ + h_\times \\[4pt] h_+ - h_\times\end{pmatrix}.
\end{equation}
Conjugating \(\mathcal C\) by \(T\) yields the diagonal operator matrix
\begin{equation}\label{eq:CT-conjugated}
T\,\mathcal C\,T^{-1} \;=\; \begin{pmatrix} C & 0\\[4pt] 0 & -C \end{pmatrix}.
\end{equation}

Because the diagonal entries are \(\pm C\), the basis \(\mathbf h_{\mathrm s}\) is an eigen-basis of the
operator matrix \(\mathcal C\) in the operator sense: the symmetric combination \(h_+^{(\mathrm s)}\) is
an eigenvector with eigen-operator \(+C\), while the antisymmetric combination \(h_-^{(\mathrm s)}\) is an
eigenvector with eigen-operator \(-C\).

Applying the same basis change to the full wave equation \eqref{eq:operator-system}
(and defining \(\mathbf S_{\mathrm s}\equiv T\mathbf S\)) produces the decoupled operator equations
\begin{equation}\label{eq:decoupled-ops}
\frac12\Box\,h_+^{(\mathrm s)} + \frac{1}{\alpha^{2}}\,C\big[h_+^{(\mathrm s)}\big] \;=\; S_+^{(\mathrm s)},\qquad
\frac12\Box\,h_-^{(\mathrm s)} - \frac{1}{\alpha^{2}}\,C\big[h_-^{(\mathrm s)}\big] \;=\; S_-^{(\mathrm s)}.
\end{equation}
These two operator equations are manifestly decoupled: each equation involves a single field and a single
action of the Cotton operator (with opposite signs for the two modes).

For completeness (and to connect with the usual circular-polarization
combinations) we recall the helicity definitions
\begin{equation}
h_R = \frac{1}{\sqrt{2}}(h_+ + i h_\times), \qquad
h_L = \frac{1}{\sqrt{2}}(h_+ - i h_\times).
\end{equation}
This transformation to the complex helicity basis is a simple unitary rotation
of the real linear polarizations, but it does \emph{not} diagonalize the Cotton
operator in the present phase convention. The operator matrix
\(\mathcal{C}=\begin{pmatrix}0 & C \\ C & 0\end{pmatrix}\) is diagonalized only
by the real symmetric/antisymmetric combinations
$(h^{(\mathrm s)}_+$ and $h^{(\mathrm s)}_-$, which correspond to eigen-operators
with eigenvalues \(+C\) and \(-C\), respectively. The helicity states
$h_R$ and $h_L$ are related to these by a further unitary rotation and therefore
mix the two eigenmodes algebraically, though the physical content is the same.
Once \(C\) is evaluated on plane-wave modes [so that it reduces to a scalar
prefactor \(f(\omega,q,k,\eta)\)], the two decoupled operator equations in the
symmetric/antisymmetric basis [Eq.~\eqref{eq:decoupled-ops}] give the scalar
wave equations with opposite-sign Cotton contributions. These are then used in
Sec.~\ref{sec:Solutions} to derive the dispersion relations.

\subsection{Source terms in the symmetric/antisymmetric basis}

For completeness, it is useful to also examine how the scalar source term
appearing in eqs.~\eqref{eq-1+}--\eqref{eq-1-} transforms under the change of basis that
diagonalizes the Cotton operator. In the $(+,\times)$ basis, both
polarizations couple to the same scalar source amplitude $S(\eta,z)$,
so that the source vector reads
\begin{equation}
\mathbf{S}(\eta,z) =
\begin{pmatrix}
S(\eta,z) \\
S(\eta,z)
\end{pmatrix}.
\end{equation}
Applying the orthogonal transformation $T$ defined in eq.~\eqref{eq:decoupled-ops}, the
symmetric/antisymmetric source vector becomes
\begin{equation}
\mathbf{S}^{(s)} = T\,\mathbf{S}
= \frac{1}{\sqrt{2}}
\begin{pmatrix}
1 & 1 \\
1 & -1
\end{pmatrix}
\begin{pmatrix}
S \\
S
\end{pmatrix}
=
\begin{pmatrix}
\sqrt{2}\,S \\
0
\end{pmatrix}.
\end{equation}
Thus, in the decoupled system only the symmetric mode $h^{(s)}_{+}$ receives a
direct driving term from the scalar background,
\begin{equation}
S_{+}(\eta,z) = \sqrt{2}\,S(\eta,z), \qquad
S_{-}(\eta,z) = 0.
\end{equation}
This shows that the antisymmetric mode evolves homogeneously in the absence
of any direct scalar forcing, while the symmetric mode carries all the
inhomogeneous contribution from the scalar background. In the helicity basis
the same result manifests as identical scalar-source amplitudes on the
right-hand side of both helicity equations.

\section{Pontryagin Constraint}
\label{sec:Pontryagin}

The Pontryagin density for the curvature is given by
\begin{equation}
    ^{\star}\mathcal{R}\mathcal{R}
    = \tensor{^{\star}R}{^{\sigma}_{\tau}^{\mu\nu}}\tensor{R}{^{\tau}_{\sigma\mu\nu}},
\end{equation}
where the relevant dual of the Riemann tensor is
\begin{equation}
    \tensor{^{\star}R}{^{\sigma}_{\tau}^{\mu\nu}}
    = \frac{1}{2}\varepsilon^{\mu\nu\alpha\beta}\tensor{R}{^{\sigma}_{\tau\alpha\beta}}.
\end{equation}
Using this definition, we get
\begin{equation}
    ^{\star}\mathcal{R}\mathcal{R} = \frac{1}{2}\varepsilon^{\mu\nu\alpha\beta}
    \tensor{R}{^{\sigma}_{\tau\alpha\beta}}\tensor{R}{^{\tau}_{\sigma\mu\nu}}.
\end{equation}
To evaluate this, we shall start by considering $\sigma = \eta$ (meaning the de Sitter time
coordinate). Then the expression takes the form
\begin{equation}
    ^{\star}\mathcal{R}\mathcal{R} \supset \frac{1}{2}\varepsilon^{\mu\nu\alpha\beta}
    \tensor{R}{^{\eta}_{\tau\alpha\beta}}\tensor{R}{^{\tau}_{\eta\mu\nu}}.
\end{equation}
The notation means that the full expression for $^{\star}\mathcal{R}\mathcal{R}$ contains terms of this
form.
Now we see that for a contribution to be potentially nonzero, $\tau$ has to be spatial---because
the only $\eta$ component in the metric is $g_{\eta\eta}$, meaning we may lower the first index
of the first Riemann tensor and make use of its antisymmetry in the first two indices when all
indices are lowered. Using this fact, we get
\begin{equation}
    ^{\star}\mathcal{R}\mathcal{R} \supset \frac{1}{2}\varepsilon^{\mu\nu\alpha\beta}
    \tensor{R}{^{\eta}_{i\alpha\beta}}\tensor{R}{^{i}_{\eta\mu\nu}}.
\end{equation}
Using all the possible permutations of the rest of the indices that we can take for the terms to be
nonzero we similarly find
\begin{equation}
    ^{\star}\mathcal{R}\mathcal{R}
    \supset \varepsilon^{\eta jzm}\tensor{R}{^{\eta}_{izm}}\tensor{R}{^{i}_{\eta\eta j}}.
\end{equation}
If we alternativel assume that $\sigma = i$ (spatial), performing the same calculations yields
\begin{equation}
     ^{\star}\mathcal{R}\mathcal{R} \supset
     \varepsilon^{\eta jzm}\tensor{R}{^{i}_{zzm}}\tensor{R}{^{z}_{i\eta j}}.
\end{equation}
So the entire form of the Pontryagin density comes out to be
\begin{equation}
     ^{\star}\mathcal{R}\mathcal{R} = \varepsilon^{\eta jzm}
     \left(\tensor{R}{^{\eta}_{izm}}\tensor{R}{^{i}_{\eta\eta j}} 
     + \tensor{R}{^{i}_{zzm}}\tensor{R}{^{z}_{i\eta j}}\right).
\end{equation}
Upon calculating the products of the Riemann tensor components involved, we get
\begin{equation}
    ^{\star}\mathcal{R}\mathcal{R} = \varepsilon^{\eta jzm}
    \left[\frac{1}{2\eta^2}\partial_{z}\partial_{\eta}h_{jm} 
    - \frac{\alpha'}{\eta^2\alpha}\partial_{z}h_{jm} 
    + \frac{1}{2}\left(\frac{\alpha'}{\alpha}\right)^2\partial_{\eta}\partial_{z}h_{mj}\right] 
    + \mathcal{O}(h^2).
\end{equation}
In the Levi-Civita tensor we may have either: $j = x$ and $m = y$, or $j = y$ and $m = x$. Using both of
these possibilities and the fact that $h_{xy} = h_{yx}$, we see that the Pontryagin constraint
is satisfied; that is, the density vanishes at first order in the perturbations:
$^{\star}\mathcal{R}\mathcal{R} = 0+\mathcal{O}(h^2)$.

\section{Form of the Scalar Field for the De Sitter Metric}
\label{sec:Scalar}

The equation of motion (\ref{eq-theta-motion}) was derived from the Lagrangian. For a massless scalar field
$\vartheta(\eta,z)$ with no self-interactions, the equation of motion takes the form
\begin{equation}
\Box \vartheta \equiv \frac{1}{\sqrt{-g}} \partial_\mu
\left(\sqrt{-g}\, g^{\mu\nu} \partial_\nu \vartheta\right) = 0.
\end{equation}

\subsection{Mode decomposition}

Writing the Fourier decomposition
\begin{equation}
\vartheta(\eta,\vec{x}\,) = \int \frac{d^3k}{(2\pi)^3} \, \vartheta_k(\eta)e^{i\vec{k}\cdot\vec{x}},
\end{equation}
the scalar wave equation reduces to an equation for each Fourier coefficient,
\begin{equation}
\vartheta_k'' + 2\frac{\alpha'}{\alpha}\,\vartheta_k' + k^2 \vartheta_k = 0,
\end{equation}
where primes denote derivatives with respect to the conformal time $\eta$. These individual Fourier modes
of $\vartheta$ represent precisely the kind of scalar we anticipated to accompany individual
propagating gravitational waves. The single-mode wave equation for a scalar
field in curved spacetime encodes how fluctuations propagate against the background geometry.
In the de Sitter case, conformal flatness allows us to recast the equation into a form resembling that
of a harmonic oscillator with a time-dependent frequency. Crucially, the competition between the
momentum term $k^2$ and the geometric contribution from the scale factor determines whether a
given mode behaves like a freely oscillating wave or whether its dynamics become frozen out.

\subsection{Rescaling}

Define the rescaled variable
\begin{equation}
\mu_k(\eta) = \alpha(\eta)\, \vartheta_k(\eta);
\end{equation}
then the equation becomes
\begin{equation}
\mu_k'' + \left(k^2 - \frac{\alpha''}{\alpha}\right)\mu_k = 0,
\end{equation}
which has the general form of the equation of motion for a parametric oscillator.
For the de Sitter metric,
\begin{equation}
\alpha(\eta) = \frac{l}{\eta}, \qquad \frac{\alpha''}{\alpha} = \frac{2}{\eta^2},
\end{equation}
and thus the mode equation reads
\begin{equation}
\mu_k'' + \left(k^2 - \frac{2}{\eta^2}\right)\mu_k = 0.
\end{equation}

\subsection{Solution}

This equation has the general solution
\begin{equation}
\mu_k(\eta) = A_k \left(1 - \frac{i}{k\eta}\right) e^{-ik\eta} + 
B_k \left(1 + \frac{i}{k\eta}\right) e^{+ik\eta}.
\end{equation}
Restoring $\vartheta_k = \mu_k / \alpha$ with $\alpha(\eta) = l/\eta$, we find
\begin{equation}
\vartheta_k(\eta) = \frac{\eta}{l} \left[
A_k \left(1 - \frac{i}{k\eta}\right) e^{-ik\eta} +
B_k \left(1 + \frac{i}{k\eta}\right) e^{+ik\eta}
\right].
\end{equation}
Imposing the Bunch-Davies vacuum condition (i.e.\ matching to flat-space positive-frequency modes
in the far past $\eta \to -\infty$) ~\cite{ref-bunchdavies} selects $B_k = 0$. The mode function then reduces to
\begin{equation}
\vartheta_k(\eta) = \frac{\eta}{l} \left(1 - \frac{i}{k\eta}\right) e^{-ik\eta}.
\end{equation}
Reassembling the Fourier expansion, a mode traveling in the $+z$-direction is
\begin{equation}
\vartheta(\eta,z) = \vartheta_{0}\frac{\eta}{l} \left(1 - \frac{i}{k\eta}\right) e^{-ik(\eta - z)}.
\end{equation}
This is the relevant
form of the solution of the massless scalar field equation $\Box\vartheta=0$ in de Sitter space.

There are two key limits to be explored. For sub-horizon fluctuations (meaning $|k\eta|\gg 1$),
$\vartheta \approx \vartheta_{0}\frac{\eta}{l}e^{-ik(\eta-z)}$, 
which is an oscillatory solution with amplitude
decaying as $1/\alpha(\eta)$.
For the super-horizon case ($|k\eta|\ll 1$), $\vartheta \approx -\vartheta_{0}\frac{i}{kl}e^{-ik(\eta-z)}$,
the field becomes frozen at a nearly constant value.

The physical behavior of the solutions depends strongly on whether one wavelength of a given mode
lies inside or outside the Hubble radius. When a mode is deep inside the horizon ($|k\eta| \gg 1$), its
wavelength is much smaller than the spacetime curvature scale, and it behaves like a standard
Minkowski-space wave, oscillating with a decaying amplitude the only evidence of the de Sitter expansion.
In contrast, once the mode crosses outside the Hubble volume ($|k\eta| \ll 1$), the expansion stretches its
wavelength beyond the causal horizon, freezing its amplitude in place. This dichotomy between sub-horizon
oscillations and super-horizon freeze-out is a cornerstone of inflationary cosmology, as it explains how
quantum fluctuations can be converted into the classical seeds of large-scale structure.

\section{Specific Solutions to the Field Equations}
\label{sec:Solutions}

\subsection{Sub-horizon limit ($|k\eta| \gg 1$)}

When we consider the sub-horizon limit, we can find the general form of the Cotton tensor elements. The
derivatives of the scalar field are 
\begin{eqnarray}
        v_{\eta} & = & \vartheta_{0}\left(\frac{1- ik\eta}{l}\right)e^{-ik(\eta - z)} \\
        v_{z} & = & \vartheta_{0}\left(\frac{ik\eta}{l}\right)e^{-ik(\eta - z)} \\
        v_{\eta\eta} & = & \vartheta_{0}\left(\frac{-k^2\eta- 2ik}{l}\right)e^{-ik(\eta - z)}\\
        v_{\eta z} & = & \vartheta_{0}\left(\frac{k^2\eta + ik}{l}\right)e^{-ik(\eta - z)}\\
        v_{zz} & = & \vartheta_{0}\left(\frac{-k^2\eta}{l}\right)e^{-ik(\eta - z)}.
\end{eqnarray}
Using these forms we get the expression for the Cotton tensor,
\begin{equation}
    C[h] = \frac{\vartheta_{0}e^{-ik(\eta - z)}}{l}\left[\left(\frac{1 - ik\eta}{2}\right)
    \partial_{z}\Box - \left(\frac{ik\eta}{2}\right)\partial_{\eta}\Box 
    + \left(-\frac{7ik}{2} + \frac{k^2\eta}{2}\right)\Box 
    - \left(ik\right)\partial_{z}\partial_{\eta}\right]h.
\end{equation}
The source term takes the form
\begin{equation}\label{eq-source}
    S(\eta,z) = 4\pi G\vartheta^{2}_{0}\left(\frac{1 - 2ik\eta}{2l^2}\right)e^{-2ik(\eta - z)}.
\end{equation}
Moreover, in the deep sub-horizon limit, the damping term in the d'Alembertian can be ignored.
It becomes negligible, and only the oscillatory terms are important. So that simplifies the form of the
Cotton tensor further, as shown in the following equations. The next step is to assume a plane wave
ansatz for each tensor mode $h_{\bullet}$, defined as
\begin{equation}
    h_{\bullet} = A_{\bullet}e^{i(\omega\eta - qz)}.
\end{equation}
Using this form and the approximations made for the $\Box$ operator, we see that
\begin{eqnarray}
    \Box h_{\bullet} & = & (-\omega^2 + q^2)h_{\bullet},\\
    \partial_{z}\Box h_{\bullet} & = & (iq\omega^2 - iq^3)h_{\bullet},\\
    \partial_{\eta}\Box h_{\bullet} & = & (-i\omega^3 + i\omega q^2)h_{\bullet}.
\end{eqnarray}
So the Cotton tensor takes the form
\begin{eqnarray}
   C[h_{\bullet}] & = & \frac{\vartheta_{0}e^{-ik(\eta - z)}}{l}
   \left[\left(\frac{1 - ik\eta}{2}\right)(-iq\omega^2 + iq^3)
   - \left(\frac{ik\eta}{2}\right)(-i\omega^3 + i\omega q^2)\right. \nonumber \\
   & & + \left. \left(-\frac{7ik}{2} + \frac{k^2\eta}{2}\right)(-\omega^2 + q^2)
   - \left(ik\right)(\omega q)\right]h_{\bullet}.
   \label{eq-pref-f}
\end{eqnarray}
 We denote this Cotton tensor prefactor by
 \begin{equation}
     C[h_{\bullet}] = f(\omega, q, k, \eta)h_{\bullet}.
 \end{equation}

The first step in finding the sub-horizon gravitational wave solutions is to examine the homogeneous
solutions of the applicable wave equations. We may then proceed to finding the solutions with
proper source terms. The complete solution will follow the standard pattern for linear systems, being
the simple sum of a general homogeneous solution and one particular sourced solution. For the homogeneous
solution we have the dispersion relations for $h^{\text{(hom)}}_{\bullet}$ set up as
\begin{equation}
(-\omega^2 + q^2)h^{(s)\text{(hom)}}_{\pm}\pm \frac{2}{\alpha^2}f(\omega, q, k, \eta)h^{(s)\text{(hom)}}_{\pm}=0.
\end{equation}
This shows that both helicities satisfy complex dispersion relations of the form
$\omega^2  = q^2 \pm (2/\alpha^{2})f(\omega, q, k, \eta)$.
Assuming that the CS correction is small and expanding around $\omega \approx q$, we get the relation
\begin{equation}
    \omega_{\pm} \approx q \pm \frac{1}{q\alpha^2}f(q, q, k, \eta).
\end{equation}
When we substitute $\omega=q$ in $f(\omega, q, k, \eta)$, all but one term drops out, leaving
the surviving leading term,
\begin{equation}
    f(q,q,k,\eta) = -\frac{i\vartheta_{0}k}{l}q^2\,e^{-ik(\eta - z)}.
\end{equation}
So the corrected frequencies in the deep sub-horizon limit become
\begin{equation}\label{eq-dis-sub}
    \omega_{\pm} \approx q\left[1 \mp \frac{ik\vartheta_{0}}{\alpha^2 l}\,e^{-ik(\eta - z)}\right].
\end{equation}

Expanding the exponential $e^{-ik(\eta - z)} = \cos\phi - i\sin\phi$, with $\phi = k(\eta - z)$, we obtain
\begin{eqnarray}
\omega_{\pm} & = & q \mp \frac{k\vartheta_{0}q}{\alpha^2 l}
\left[\sin\phi + i\cos\phi\right].
\end{eqnarray}
Thus, the correction has both \emph{real} and \emph{imaginary} parts:
\begin{equation}\label{eq-dis-sub-separated}
    \Re\{\omega_{\pm}\} = q \mp \frac{k\vartheta_{0}q}{\alpha^2 l}\sin\phi, \qquad
    \Im\{\omega_{\pm}\} = \mp \frac{k\vartheta_{0}q}{\alpha^2 l}\cos\phi.
\end{equation}
The real part introduces a small phase-velocity shift, corresponding to \emph{velocity (phase) birefringence}, 
while the imaginary part introduces a helicity-dependent damping or amplification, corresponding to
\emph{amplitude birefringence}.
Depending on the phase $\phi$ of the background scalar field, one of these effects dominates.
\begin{itemize}
    \item For $\phi \approx 0$ (scalar and gravitational wave in phase), $\cos\phi \approx 1$, and the correction
    is mostly imaginary---producing amplitude birefringence.
    \item For $\phi \approx \pi/2$, $\sin\phi \approx 1$, the correction is mostly real---producing
    velocity birefringence.
    \item For general $\phi$, both amplitude and velocity birefringence coexist.
\end{itemize}
Physically, this means that the CS background alternately modulates the amplitude and phase of each
helicity as the scalar field and the gravitational wave move through each other, transferring energy between the
two helicities depending on their relative phase.

In terms of conformal time, $\alpha(\eta) = l/\eta$ implies $1/(\alpha^{2}l) = \eta^2/l^3$, so the\
envelope of both effects grows
as $\eta^2$ in the sub-horizon regime, while oscillating with $\phi=k(\eta-z)$. Since $\eta=-e^{-Ht}/H$,
this means both
phase and amplitude birefringence are strongest in the early universe ($|\eta|\gg1$) and decay exponentially
in cosmic time.

It is important to emphasize the different roles played by the two wave numbers that appear in our formulas.
The symbol $q$ denotes the spatial momentum of the \emph{gravitational wave perturbation}, which enters
through the plane-wave ansatz
$h \sim e^{i(\omega\eta - qz)}$. In contrast, $k$ is the wave number associated with the
\emph{Chern-Simons scalar background} $\vartheta(\eta,z)$, which we have taken to
oscillate as $\vartheta \sim \vartheta_0 f(\eta)\,e^{-ik(\eta-z)}$. Both $q$ and $k$ therefore appear in
the Cotton tensor, because derivatives act on both the tensor mode
and the scalar background simultaneously. Physically, the two wave numbers need not be equal; a gravitational
wave with wavelength $2\pi/q$ may propagate in a background
scalar field oscillating with a different wavelength $2\pi/k$. Only in special limits, such as a
spatially homogeneous scalar background ($k=0$), would one of these drop out.
Throughout this section we therefore keep $q$ and $k$ distinct, with $q$ governing the propagation of the
tensor perturbations and $k$ setting the modulation scale of the
parity-violating background. The simultaneous presence of both real and imaginary parts in
eq.~\eqref{eq-dis-sub} shows that the birefringence manifests as a combined
\emph{phase and amplitude} effect, whose relative strength oscillates with the scalar phase $\phi=k(\eta-z)$.

\subsection{Particular solution}

The next step is to find a single inhomogeneous (or particular) solution for the
equation including the source term. Since the source is proportional to
$e^{-2ik(\eta - z)}$ [see eq.~\eqref{eq-source}], we assume the same functional form for the
particular solution. However, because in the symmetric/antisymmetric basis only
the symmetric mode $h^{(s)}_{+}$ is sourced (cf.\ sec.~\ref{sec:diagonalization}), the Ansatz applies
only to $h^{(s)}_{+}$:
\begin{equation}
    h^{(s)\,\text{(part)}}_{+} = A_{+}(\eta)\,e^{-2ik(\eta - z)}.
    \label{eq-part-ansatz}
\end{equation}
Here $A_{+}(\eta)$ is a slowly varying amplitude that can be treated as nearly
constant in the deep sub-horizon limit. Applying $\Box$ on this form of the
Ansatz shows that the leading oscillatory pieces cancel, as the exponential is
a solution of the free wave equation. Physically, this reflects that the
contributing sources lie on the past light cone. The surviving contribution
comes from the Cotton tensor terms. Algebraically we obtain
\begin{equation}
    \left[\frac{2}{\alpha^2}f(2k, 2k, k, \eta)\right]\,h^{(s)\,\text{(part)}}_{+}
    = \sqrt{2}\,S(\eta,z)
    = 4\pi G\vartheta_{0}^{2}\left(\frac{1 - 2ik\eta}{\sqrt{2}l^2}\right)e^{-2ik(\eta - z)}.
\end{equation}
So the particular solution is
\begin{equation}
    h^{(s)\,\text{(part)}}_{+}
    = 4\pi G\vartheta_{0}^{2}\left(\frac{1 - 2ik\eta}{\sqrt{2}l^2}\right)
      \frac{e^{-2ik(\eta - z)}}{\tfrac{2}{\alpha^2}f(2k,2k,k,\eta)}.
\end{equation}

The Cotton prefactor at $(\omega,q)=(2k,2k)$ simplifies to
\begin{equation}
    f(2k,2k,k,\eta) \;\approx\; -\frac{4i\vartheta_{0}k^3}{l}\,e^{-ik(\eta-z)}.
\end{equation}
Plugging this back in, the particular solution reduces to
\begin{equation}
    h^{(s)\,\text{(part)}}_{+}
    = \frac{i\pi G \vartheta_{0} l}{2\sqrt{2}\eta^2k^3}\,(1 - 2ik\eta)\,e^{-ik(\eta - z)}.
    \label{eq-part-sol}
\end{equation}
In the deep sub-horizon regime $|k\eta|\gg1$ this further simplifies to
\begin{equation}
    h^{(s)\,\text{(part)}}_{+}
    \approx \,\frac{\pi G \vartheta_{0} l}{\sqrt{2}k^2\eta}\,e^{-ik(\eta - z)}.
\end{equation}

Thus the particular solution is another oscillatory wave at frequency $k$
sourced by the scalar field background. Its amplitude decays as $1/\eta$ in
conformal time, corresponding to an exponential growth $\sim e^{Ht}$ in cosmic
time. Although at first sight such growth may appear unphysical for
gravitational waves, it is expected here; the inhomogeneous solution is
continuously driven by the scalar field source through the CS coupling.
Physically, the scalar background acts as a persistent pump field that transfers
energy into the tensor sector, leading to amplification of the symmetric mode.
This growth therefore does not indicate a pathology but is a known feature of
CS-modified gravity. In realistic scenarios the amplification would eventually
be regulated by back-reaction once the tensor modes become strong enough to
drain energy from the scalar background.

It is also important to note that in the deep sub-horizon regime ($|k\eta|\gg
1$), the particular solution is suppressed by the prefactor $\sim(k^2\eta)^{-1}$,
so for large $|k\eta|$ the sourced contribution is negligible compared to the
homogeneous oscillatory modes. Only as the mode approaches the horizon
($|k\eta|\sim 1$) does this suppression weaken, and in the super-horizon regime
($|k\eta|\ll 1$) the amplitude grows, reflecting the continuous pumping of tensor
modes by the scalar background through the CS coupling.

\subsection{Super-horizon limit ($|k\eta| \ll 1$)}

Turning to the super-horizon limit ($|k\eta| << 1$), we have a form for the field that is frozen with a
constant amplitude,
\begin{equation}
    \vartheta \approx -\frac{i\vartheta_{0}}{kl}e^{-ik(\eta - z)}.
\end{equation}
In this limit we can once again get the Cotton tensor elements, starting with the first and second
derivatives of the scalar,
\begin{eqnarray}
         v_{\eta} & = & -\frac{\vartheta_{0}}{l}e^{-ik(\eta - z)} \\
         v_{z} & = & \frac{\vartheta_{0}}{l}e^{-ik(\eta - z)}\\
        v_{\eta\eta} & = & \frac{ik\vartheta_{0}}{l}e^{-ik(\eta - z)}\\
        v_{\eta z} & = & -\frac{i\vartheta_{0}k}{l}e^{-ik(\eta - z)}\\
        v_{zz} & = & \frac{i\vartheta_{0}k}{l}e^{-ik(\eta - z)}.
\end{eqnarray}
Using these, we get that the Cotton tensor in the presence of the tensor fluctuations takes the form
\begin{equation}
    C[h_{\bullet}] = \frac{\vartheta_{0}e^{-ik(\eta - z)}}{l}\left[-\frac{1}{2}\partial_{z}\Box
    - \frac{1}{2}\partial_{\eta}\Box - \left(\frac{4}{\eta} + \frac{ik}{2}\right)\Box\right]h_{\bullet}.
\end{equation}
However, quite importantly, in this limit the source term $S(\eta, z)$ vanishes identically at leading order,
$S(\eta,z)=0$. There will consequently be no particular solution to find.

However, there is a different source of complexity in the solutions. In the super-horizon limit we
cannot ignore the damping term that arises from the action of the d'Alembertian; the damping appears at
leading order. So again assuming a plane wave Ansatz, we get
\begin{eqnarray}
        \Box h_{\bullet} & = & \left(-\omega^2 + q^2 - \frac{2i\omega}{\eta}\right)h_{\bullet} \\
    \partial_{z}\Box h_{\bullet} & = & \left(iq\omega^2 - iq^3 - \frac{2q\omega}{\eta}\right)h_{\bullet} \\
    \partial_{\eta}\Box h_{\bullet} & = & \left(-i\omega^3 + i\omega q^2 + \frac{2\omega^2}{\eta}
    + \frac{2i\omega}{\eta^2}\right)h_{\bullet}.
\end{eqnarray}
Plugging these into the Cotton tensor term and extracting the prefactor, we get another function, of the form
\begin{equation}
    C[h_{\bullet}] = g(\omega, q, k, \eta)h_{\bullet},
\end{equation}
where
\begin{eqnarray}
    g(\omega, q, k, \eta)=\frac{\vartheta_{0}e^{-ik(\eta - z)}}{l}\!\!\!\!\!\!\!\!\!\! & &
    \left[-\frac{iq\omega^2}{2}
    + \frac{iq^3}{2} + \frac{\omega q}{\eta} + \frac{i\omega^3}{2} - \frac{i\omega q^2}{2}
    + \frac{3\omega^2}{\eta}\right. \nonumber\\
    & & + \left.\frac{7i\omega}{\eta^2} - \frac{4q^2}{\eta} + \frac{ik\omega^2}{2} - \frac{ikq^2}{2}
    - \frac{\omega k}{\eta}\right]\!\! ;
\end{eqnarray}
and using this explicit form in the wave equations we get the dispersion relation,
\begin{equation}
    \omega^{2}_{\pm} = q^2 - \frac{2i\omega}{\eta} \pm \frac{2}{\alpha^2}g(\omega, q, k, \eta).
\end{equation}

Again perturbing around $\omega \approx q$, we get a simplified version of the prefactor,
\begin{equation}
    g(q, q, k, \eta) = \left[\frac{7iq}{\eta^2} 
    - \frac{qk}{\eta}\right]\frac{\vartheta_{0}e^{-ik(\eta - z)}}{l}
     = \frac{q}{\eta^2}\left(7i - k\eta\right)\frac{\vartheta_{0}e^{-ik(\eta - z)}}{l}.
\end{equation}
In the super-horizon limit we have $|k\eta| \ll 1$, so $k\eta$ may be ignored compared with $7i$,
and we get a finite birefringence independent of $\eta$, as follows
\begin{equation}
    \omega^{2}_{\pm} = q^2 - \frac{2i\omega}{\eta} \pm i\frac{14q\vartheta_{0}}{l^3}e^{-ik(\eta - z)}.
\end{equation}
Conveniently, we can get a clean expression the frequency itself (rather than its square),
\begin{equation}
    \omega_{\pm} \approx q - \frac{i}{\eta} \pm  i\frac{7\vartheta_{0}}{l^3}e^{-ik(\eta - z)}.
\end{equation}
Again on expanding the exponential $e^{-ik(\eta - z)} = \cos\phi - i\sin\phi$ with $\phi = k(\eta - z)$
we get a real and imaginary part for the correction as:
\begin{equation}
    \Re\{\omega_{\pm}\} = q \pm \frac{7\vartheta_{0}}{l^3}\sin\phi,\qquad
    \Im\{\omega_{\pm})\} = -\frac{1}{\eta} \pm \frac{7\vartheta_{0}}{l^3}\cos\phi.
\end{equation}
The term $-1/\eta$ encodes the usual Hubble friction, which damps both helicities equally,
while the $\cos\phi$ and $\sin\phi$ components produce helicity-dependent amplification and phase shifts.

The point to note here is that the amplitude of the correction terms are independent of $\eta$. Physically,
this means that even after horizon crossing, when the scalar field becomes nearly frozen,
the CS term continues to imprint a residual parity-violating birefringence. The real component
slightly shifts the propagation phase between left and right helicities (a velocity birefringence),
while the imaginary component introduces differential damping (amplitude birefringence). As the
universe expands ($\eta \rightarrow 0^{-}$), both helicities undergo the common exponential
damping $e^{-Ht}$ due to Hubble friction, while the relative birefringence imprint remains as a
frozen parity-violating signature in the primordial tensor spectrum.

\subsection{Birefringence effects in sub- and super-horizon regimes}

In the sub-horizon regime ($|k\eta|\gg 1$), the dispersion relation takes the form
\begin{equation}
\omega_\pm \approx q \mp i\frac{k\vartheta_{0}q}{\alpha^2 l}\, e^{-ik(\eta-z)},
\end{equation}
so that the birefringence (frequency splitting between the two helicities) is
\begin{equation}
\Delta\omega \equiv \omega_+ - \omega_- \approx -\,i\frac{2k\vartheta_{0}q}{\alpha^2 l}\, e^{-ik(\eta-z)}.
\label{eq-sub-Delta}
\end{equation}
Using $\alpha(\eta)=l/\eta$, this becomes
\begin{equation}
\Delta\omega = -i\frac{2k\vartheta_{0}q\eta^2}{l^3}e^{-ik(\eta - z)}.
\end{equation}
Thus the birefringence amplitude grows quadratically with conformal time.
\begin{figure}[h]
    \centering
    \includegraphics[width=0.5\linewidth]{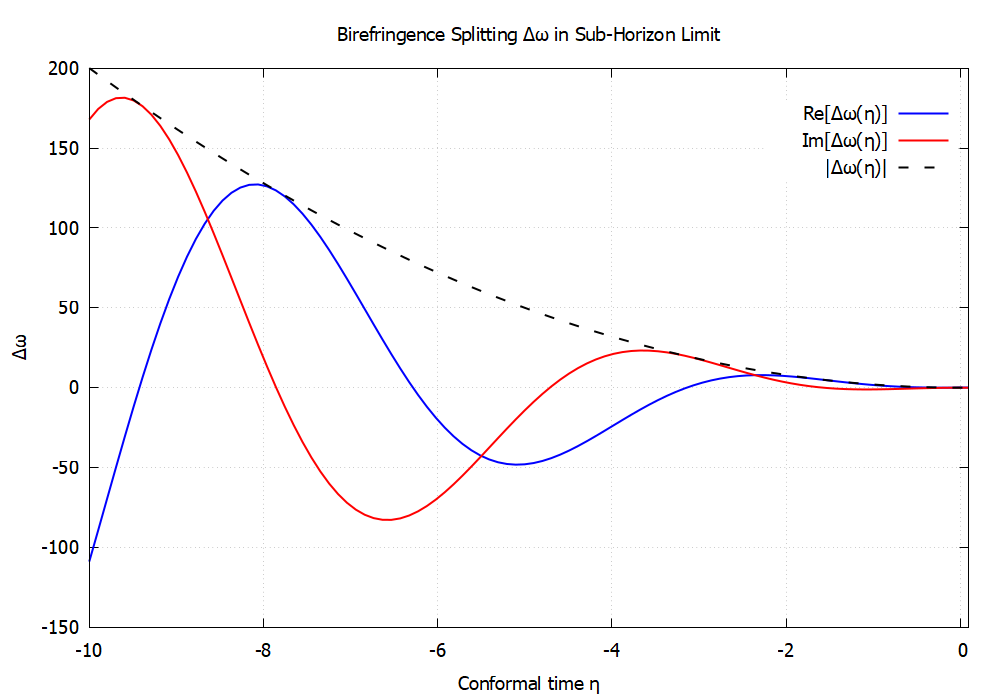}
    \caption{The birefringence effect ($\Delta\omega$) versus conformal time ($\eta$) in the
    sub-horizon limit. The red line shows the imaginary part, and the blue line shows the real part
    for the calculated birefringence. The dashed line is the absolute value for $\Delta\omega$,
    which envelops the real and imaginary parts.}
    \label{fig:bire_sub_horizon}
\end{figure}

In the sub-horizon regime, the birefringence splitting $\Delta \omega(\eta)$ exhibits a rich
oscillatory structure, with real and imaginary parts alternating in dominance as conformal
time evolves. The real component corresponds to phase differences between the two helicities,
while the imaginary component encodes the Hubble damping effect on their amplitudes. As
shown in fig.~\ref{fig:bire_sub_horizon}, the overall magnitude $|\Delta \omega|$ decays with time,
indicating that the parity-violating birefringence is strongest in the early universe.
This behavior highlights the transient yet potentially
observable nature of CS-induced chiral effect in the primordial gravitational
wave spectrum.

On the other hand, in the super-horizon regime ($|k\eta|\ll 1$) where the scalar source vanishes, the birefringence splitting is given by
\begin{equation}
\Delta\omega = \omega_+ - \omega_-\approx i\frac{14\vartheta_{0}}{l^3} e^{-ik(\eta-z)},
\label{eq-super-Delta}
\end{equation}
for which the magnitude is independent of $\eta$. This shows that while both helicities
are damped by expansion, the relative splitting remains constant in amplitude. In cosmic time,
this birefringence survives as a frozen parity-violating imprint in the super-horizon regime.

\begin{figure}[h]
    \centering
    \includegraphics[width=0.5\linewidth]{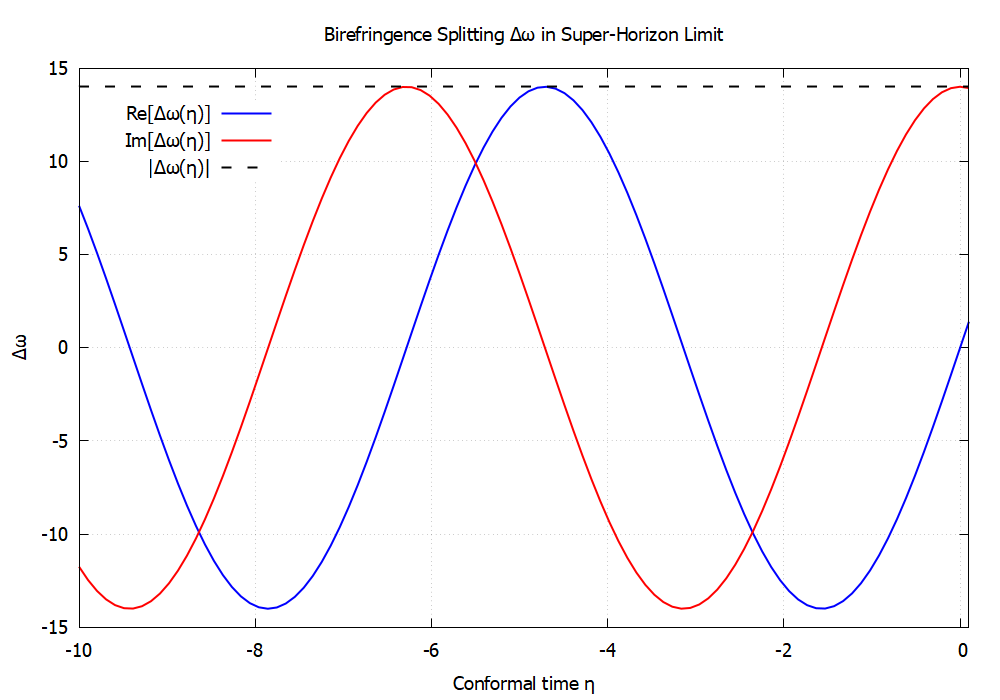}
    \caption{The amplitude birefringence $\Delta\omega$ versus conformal time in the super-horizon limit.
    The red line again shows the imaginary part and the blue the real part, with the dashed line indicating
    the constant total magnitude.}
    \label{fig:bire_super_horizon}
\end{figure}

In the super-horizon regime, the birefringence splitting $\Delta \omega(\eta)$ approaches a
frozen amplitude, as seen from the constant $|\Delta \omega|$. While the real
and imaginary parts oscillate with conformal time, their envelope remains bounded, reflecting
the fact that Hubble friction damps the overall growth of tensor perturbations. Unlike the
sub-horizon case, where birefringence decays away with time, the super-horizon splitting
stabilizes to a constant value that is independent of $\eta$. This frozen birefringence imprint
persists even as the modes are stretched far beyond the causal horizon, potentially leaving behind a
permanent parity-violating signature in the primordial gravitational wave spectrum.

So in the sub-horizon regime, the birefringence rate oscillates with the scalar phase and decays
exponentially in cosmic time. Once the super-horizon regime is reached, the birefringence amplitude
becomes constant (independent of $\eta$), while both helicities undergo exponential Hubble damping.
This distinction highlights that CS-induced parity violation is an early-universe effect that
nonetheless may leave persistent imprints on super-horizon tensor modes while becoming
negligible for sub-horizon modes at late times.

\section{Particular Solution and Flat-Spacetime Limit}
\label{sec:Flux}

In the presence of a scalar background $\vartheta(\eta,z)$, the inhomogeneous part of the
gravitational wave equation admits a nontrivial particular solution. From our analysis [see
eqs.~(\ref{eq-part-ansatz}--\ref{eq-part-sol})], the sub-horizon particular solution
(the sourced helicity) takes the form
\begin{equation}
    h^{(s)(\text{part})}_{+}(\eta,z) \;\approx\; 
     \frac{\pi G \,\vartheta_{0}\, l}{\sqrt{2}\,k^2 \,\eta}\, e^{-ik(\eta-z)} ,
    \label{eq:hpart_subhorizon_corrected}
\end{equation}
while the $h_{-}$ mode has no inhomogeneous source and therefore contains only the homogeneous solution.
(Hence we take its sourced amplitude to vanish for the particular-solution flux computation.)
The particular solution decays with conformal time as \(1/\eta\), reflecting the redshifting of the 
source in the expanding de~Sitter background; it is also suppressed by \(1/k^2\), consistent with the
intuitive expectation that short-wavelength modes are less efficiently sourced.

\subsection{Flux in the sub-horizon regime}

The Isaacson energy flux ~\cite{ref-isaacson1, ref-isaacson2} along the propagation direction (with $\omega\approx k$) is
\begin{equation}
    \mathcal{F} \;=\; \frac{\omega^2}{32\pi G}\left[|h^{(s)}_{+}|^2 + |h^{(s)}_{-}|^2\right]
    \;\approx\; \frac{k^2}{32\pi G}\,|h^{(s)}_{+}|^2,
\end{equation}
where in the last step we have used that the sourced contribution comes only from $h^{(s)}_{+}$.
Using (\ref{eq:hpart_subhorizon_corrected}) we obtain
\begin{equation}
    |h^{(s)}_{+}|^2 \;=\; \frac{\pi^2 G^2 \vartheta_{0}^2 l^2}{2\,k^4 \eta^2},
\end{equation}
so the flux contributed by the particular solution is
\begin{equation}
    \mathcal{F}
    \;=\; \frac{k^2}{32\pi G}\,\frac{\pi^2 G^2 \vartheta_{0}^2 l^2}{2\,k^4 \eta^2}
    \;=\; \frac{\pi G}{64}\,\frac{l^2}{k^2 \eta^2}\,\vartheta_{0}^2.
    \label{eq:F_subhorizon_corrected}
\end{equation}

This expression exhibits the expected scalings:
\begin{itemize}
    \item $\mathcal{F}\propto \vartheta_{0}^2$: There is a quadratic dependence on the scalar amplitude.
    \item $\mathcal{F}\propto 1/k^2$: Short wavelength modes radiate less efficiently.
    \item $\mathcal{F}\propto 1/\eta^2$: The particular-solution flux grows as conformal time approaches \(0^-\),
          equivalently as cosmic time increases (because the sourced amplitude redshifts like \(1/\eta\)).
    \item $\mathcal{F}\propto l^2$: The expression has an explicit dependence on the de~Sitter radius.
\end{itemize}

\subsection{Flat-spacetime limit}

The Minkowski (flat-spacetime) limit corresponds to the scale factor approaching unity,
\(\alpha(\eta)=\tfrac{l}{\eta}\to 1\) (equivalently \(H\to 0\)). It is therefore appropriate to take the
limit \(\tfrac{l}{\eta}\to 1\) in the expression for the amplitude rather than setting \(\eta=l\) as
both may vanish. In this limit the particular-solution amplitude becomes
\begin{equation}
    h^{(s)(\text{part})}_{+}\big|_{\alpha\to 1}\;\approx\;
    \frac{\pi G \,\vartheta_{0}}{\sqrt{2}\,k^2}\,e^{-ik(\eta-z)} ,
    \label{eq:hpart_flat_limit}
\end{equation}
and $h_{-}$ remains unsourced. (Only its homogeneous piece survives.) The corresponding flux in the
flat limit is given by setting \(l/\eta\to 1\) in (\ref{eq:F_subhorizon_corrected}),
\begin{equation}
    \mathcal{F}_{\text{Minkowski}} \;=\; \frac{\pi G}{64}\,\frac{1}{k^2}\,\vartheta_{0}^2.
\end{equation}
Thus the de~Sitter particular-solution result reduces smoothly to the expected Minkowski scaling
\(\mathcal{F}\propto \vartheta_{0}^2/k^2\) [with the numerical coefficient above determined by our
conventions and the \(\sqrt{2}\) normalization used in \eqref{eq:hpart_subhorizon_corrected}].

\subsection{Flux in the super-horizon regime}

Outside the horizon the source vanishes, \(S(\eta,z)=0\), so no net pumping of energy into tensor modes
occurs. The tensor modes evolve according to homogeneous evolution and are damped by expansion:
\begin{equation}
    h^{(s)}_\pm \sim e^{-Ht}=e^{-t/l},
\end{equation}
so the Isaacson flux (which scales as $|\partial_t h|^2\propto h^2$ for these modes) decays as
\begin{equation}
    \mathcal{F}\propto e^{-2Ht}.
\end{equation}
At late times the radiated flux therefore vanishes, although a frozen birefringent frequency splitting
can remain imprinted in the primordial tensor spectrum. The key point for the present paper is that the
sourced radiative flux originates only from the \(h^{(s)}_{+}\) helicity in our setup;
the \(h^{(s)}_{-}\) helicity
receives no source contribution and only carries whatever homogeneous (unsourced) amplitude was present.

\section{Phase Difference Between Helicity States}
\label{sec:Phase}

A key observable manifestation of birefringence is the accumulated phase difference between
the right- and left-handed helicity states of the tensor perturbations. It quantifies the relative
propagation speed of the two circular polarizations and is defined as
\begin{equation}\label{eq-delta_phi}
\Delta\phi(k, \eta) = \int^{\eta} d\eta' \, \Re\{\Delta\omega(\eta')\},
\end{equation}
where $\Delta\omega(\eta) = \omega_{+} - \omega_{-}$ is the instantaneous frequency splitting
between the two helicities. As discussed in sec.~\ref{sec:Solutions}, this splitting is a complex, oscillatory
quantity whose real part induces velocity (phase) birefringence, while the imaginary part
controls helicity-dependent amplitude modulation.

It is important to specify the lower integration limit in eq.~\eqref{eq-delta_phi}. Physically, the phase
difference should begin accumulating only after the tensor modes become well-defined,
coherent oscillations. For sub-horizon modes ($|k\eta|\gg 1$), this condition is satisfied
in the far past, when the modes behave as free plane waves in the Bunch-Davies vacuum;
hence we choose $\eta_{\mathrm{in}}\to -\infty$, ensuring that early-time oscillations contribute
a vanishing net phase shift. In contrast, for super-horizon modes ($|k\eta|\ll 1$), the
wave behavior ceases once the mode crosses the horizon. The relevant lower limit is then the
horizon-crossing time $\eta_{\mathrm{in}}=-1/k$, when the wavelength first equals the Hubble
radius. Earlier contributions are exponentially suppressed by Hubble damping and do not
affect the observable phase offset. Accordingly, we write
\begin{equation}
\Delta\phi(k, \eta) = \int_{\eta_{\mathrm{in}}}^{\eta}d\eta'\, \Delta\omega(\eta'),
\end{equation}
with $\eta_{\mathrm{in}}$ chosen as $-\infty$ for sub-horizon and $-1/k$ for super-horizon analyses.

\subsection{Sub-horizon regime}

Using the dispersion relation (\ref{eq-dis-sub}) valid for $|k\eta|\gg1$, 
the instantaneous frequency splitting is
\begin{equation}
\Re\{\Delta\omega(\eta)\} = -\frac{2k\vartheta_0 q}{l^3}\,\eta^2 \sin{k(\eta - z)} .
\end{equation}
Integrating gives
\begin{equation}
    \Delta\phi_{\text{sub}}(k,\eta)
    = -\frac{2k \vartheta_0 q}{l^3}
      \int_{\eta_{\rm in}}^{\eta} d\eta'\,\eta'^2 \sin{k(\eta' - z)} .
\end{equation}
Here $\eta_{\rm in}$ denotes the time when the mode is initialized in the
Bunch--Davies vacuum.  
For calculations one may keep $\eta_{\rm in}$ finite (e.g.\ when a source turns on)
or take the formal limit $\eta_{\rm in}\!\to\!-\infty$ with the usual $i\epsilon$ Tauberian regulator, 
which suppresses early oscillations.

Carrying out the integration explicitly yields
\begin{align}
    I(\eta) = \int_{\eta_{\rm in}}^{\eta}d\eta'\, \eta'^2 \sin{k(\eta' - z)}
    &= \left(-\frac{\eta^2}{k} + \frac{2}{k^3}\right)\cos{k(\eta - z)} + \frac{2\eta}{k^2}\sin{k(\eta - z)} +
    \nonumber\\ &\left(\frac{\eta^2_{\rm in}}{k} - \frac{2}{k^3}\right)\cos{k(\eta_{\rm in} - z)} -
    \frac{2\eta_{\rm in}}{k^2}\sin{k(\eta_{\rm in} - z)}.
\end{align}
In the limit $\eta_{\rm in}\!\to\!-\infty$ the regulated second term vanishes, leaving
\begin{equation}
    I(\eta) = \left(-\frac{\eta^2}{k} + \frac{2}{k^3}\right)\cos{k(\eta - z)}+\frac{2\eta}{k^2}\sin{k(\eta - z)}.
\end{equation}
For $|k\eta|\gg1$ one may expand the expression as:
\begin{eqnarray}
    I(\eta) = \frac{\eta^2}{k}\left\{-\cos{[k(\eta - z)]} + \frac{2}{k\eta}\sin{[k(\eta - z)]}+
    \frac{2}{(k\eta)^2}\cos{[k(\eta - z)]} + \mathcal{O}[(k\eta)^{-3}]\right\}.
\end{eqnarray}

The dominant contribution for large $|k\eta|$ is therefore
\begin{equation}
    \Delta\phi_{\text{sub}}(k,\eta)\big|_{\text{leading}}
    = \frac{2\vartheta_0 q}{l^3}\,\eta^2\,\cos{[k(\eta - z)]},
\end{equation}
showing that the envelope of the accumulated phase difference grows as $\eta^2$,
which corresponds to an exponential decay $\propto e^{-2Ht}$ in cosmic time 
($\eta=-e^{-Ht}/H$), since $|\eta|$ decreases exponentially with $t$.

\subsection{Super-horizon regime}

In the super-horizon limit ($|k\eta|\ll1$), the real part of the frequency splitting between the two
helicities is
\begin{equation}
   \Re\{\Delta\omega(\eta)\}
   = \frac{14\vartheta_0}{l^3}\sin\left[k(\eta-z)\right].
\end{equation}
The accumulated phase difference, obtained by integrating this real part from the horizon-crossing
time $\eta_{\rm in}=-1/k$ to $\eta$, is
\begin{equation}
   \Delta\phi_{\text{super}}(k,\eta)
   = \frac{14\vartheta_0}{l^3}\frac{1}{k}
     \left[\cos k(\eta_{\rm in}-z) - \cos k(\eta-z)\right].
   \label{eq:super-phase-exact}
\end{equation}
This is the exact expression for the accumulated phase in the super-horizon regime.

For modes well outside the horizon, $|k\eta|\ll1$, the cosine terms in
eq.~\eqref{eq:super-phase-exact} may be expanded as
\begin{equation}
   \cos k(\eta-z)\simeq \cos(-kz)\!\left[1-\frac{(k\eta)^2}{2}\right]
           -\sin(-kz)\,k\eta.
\end{equation}
Taking the difference at $\eta$ and $\eta_{{\rm in}}$ and simplifying gives
\begin{equation}
   \cos k(\eta_{\rm in}-z)-\cos k(\eta-z)
   \simeq
   \frac{k^2}{2}\cos(kz)\,(\eta^2-\eta_{\rm in}^2)
   + k\sin(kz)\,(\eta_{\rm in}-\eta).
\end{equation}
Substituting this into Eq.~\eqref{eq:super-phase-exact} yields
\begin{equation}
   \Delta\phi_{\text{super}}(k,\eta)
   \simeq
   \frac{14\vartheta_0}{l^3}
   \left[
      \frac{k}{2}\cos(kz)\,(\eta^2-\eta_{\rm in}^2)
      + \sin(kz)\,(\eta_{\rm in}-\eta)
   \right].
   \label{eq:super-phase-expanded}
\end{equation}

The first term in eq.~\eqref{eq:super-phase-expanded} is $\mathcal{O}(k)$
and thus subleading; the second term dominates for small~$k$.
Retaining only the leading-order contribution gives
\begin{equation}
   \Delta\phi_{\text{super}}(k,\eta)\big|_{\text{leading}}
   \approx
   \frac{14\vartheta_0}{l^3}\sin(kz)\,(\eta_{\rm in}-\eta).
\end{equation}
With $\eta_{\rm in}=-1/k$, this becomes
\begin{equation}
   \Delta\phi_{\text{super}}(k,\eta)
   \approx
   -\frac{14\vartheta_0}{k\,l^3}\sin(kz)
   -\frac{14\vartheta_0}{l^3}\sin(kz)\,\eta.
\end{equation}
Since $|\eta|\ll1/k$ for super-horizon modes, the $\mathcal{O}(1/k)$ term
dominates.  The accumulated phase therefore saturates to a constant value after horizon exit,
representing a frozen, parity-violating offset imprinted on the tensor modes.

\begin{figure}[h]
    \centering
    \includegraphics[width=0.75\linewidth]{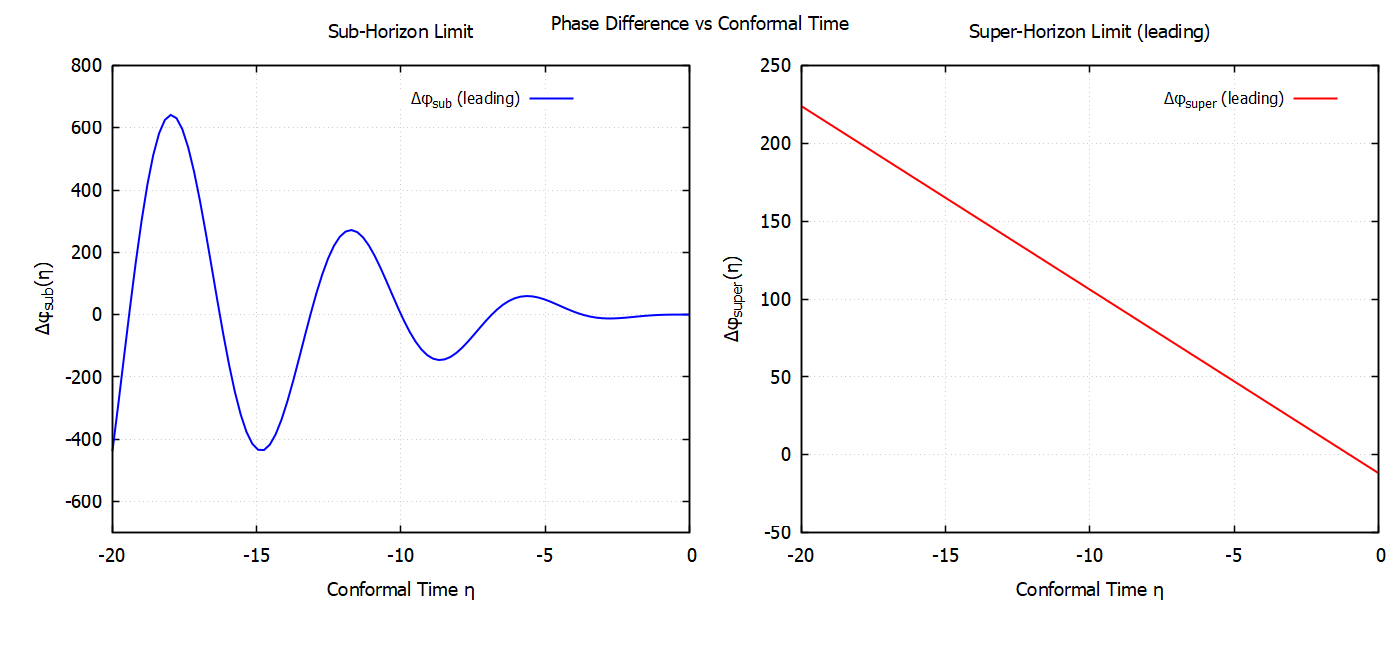}
    \caption{The leading order phase differences ($\Delta\phi$) versus conformal time for
    both the sub- and super-horizon limits (in arbitrary units).
    The plot on left for the sub-horizon limit shows the quadratic
    dependence ($\propto \eta^2 $). The plot on right for the super-horizon limit shows the decaying
    parts as conformal time reaches the far future
    ($\eta \rightarrow 0^{-}$).}
    \label{fig:phase_diff}
\end{figure}

Figure~\ref{fig:phase_diff} illustrates the phase difference $\Delta \phi$ between the helicity states
of tensor perturbations as a function of conformal time $\eta$ in both the sub-horizon (left) and
super-horizon (right) limits. In the sub-horizon regime, the phase difference grows quadratically with
$\eta$, reflecting the scaling $\Delta \phi_{\text{sub}} \propto \eta^{2}$. This behavior is consistent
with the analytic result that birefringence effects are strongest at early conformal times and decay
exponentially in cosmic time as the universe expands. In contrast, the super-horizon regime shows a linear
dependence on $\eta$, modulated by oscillatory factors from the scalar background giving rise to a frozen
parity-violating phase difference persists once the modes exit the horizon. This distinction highlights how
sub- and super-horizon dynamics imprint qualitatively different signatures on the primordial gravitational
wave spectrum.

\section{Amplitude Birefringence}

The amplitude ratio $A_+/A_-$ is another important observable that quantifies the differential amplification
or attenuation of the two polarization states of gravitational waves induced by the CS coupling.
It is calculated using the expression,
\begin{equation}
    \frac{A_{+}}{A_{-}} = \text{exp}\left[\int^{\eta}_{\eta_{\rm in}}d\eta'\,
    \Im\{\Delta\omega(\eta')\}\right].
\end{equation}
Since the imaginary part of the modified dispersion relation enters the wave amplitude as an exponential
factor, integrating $\Im\{\Delta\omega\}$
directly measures the net birefringent amplification accumulated during propagation. 
This ratio therefore serves as a gauge of \emph{amplitude birefringence}; a nonzero value signals that
parity violation in the CS term has led to unequal damping or enhancement of the left- and right-handed
gravitational wave modes. 
In cosmological settings, this effect provides an observational handle to probe parity-violating interactions
in the early universe and can, in principle, imprint a helicity-dependent modulation in the stochastic
gravitational wave background~\cite{ref-wen}.

\subsection{Sub-horizon regime}

Inside the sub-horizon regime ($|k\eta|\gg1$), the CS modification introduces a small imaginary component to
the frequency of each helicity mode. 
From eqs.~\eqref{eq-dis-sub}--\eqref{eq-dis-sub-separated}, the imaginary parts of the two helicity branches are
\begin{equation}\label{eq:amp_ratio_def}
    \Im\{\omega_\pm\} = \mp\,\frac{k\vartheta_0\,q}{\alpha^2 l}\,\cos{[k(\eta - z)]},
    \end{equation}
where $\vartheta_0$ characterizes the CS coupling amplitude, and $\alpha(\eta) = l / \eta$ for the de
Sitter background. 
The imaginary part of the helicity frequency splitting $\Delta\omega \equiv \omega_+ - \omega_-$ is then
\begin{equation}
    \Im\{\Delta\omega\} = \Im\{\omega_+\}- \Im\{\omega_-\}
    = -\frac{2k\vartheta_0\,q\,\eta^2}{l^3}\cos{[k(\eta - z)]}.
    \label{eq:Im_Domega}
\end{equation}

The relative amplitude of the two helicity states follows from integrating the imaginary frequency difference.
To evaluate the exponent, define
\begin{equation}
    I'(\eta) \equiv 
    \int^{\eta} d\eta'\, \eta'^2 \cos\left[k(\eta' - z)\right].
\end{equation}
An explicit anti-derivative is given as:
\begin{equation}
    I'(\eta) = 
    \frac{
    \eta^2 k^2 \sin[k(\eta - z)] 
    + 2\eta k \cos[k(\eta - z)] 
    - 2\sin[k(\eta - z)]
    }{k^3} + \text{const.}
\end{equation}
Substituting this into Eq.~\eqref{eq:amp_ratio_def}, we obtain the exact sub-horizon expression
\begin{align}
    \ln\!\frac{A_+}{A_-}
    &= 
    -\,\frac{2 k\vartheta_0 q}{l^3}
    \Bigg\{
    \frac{
    \eta^2 k^2 \sin[k(\eta - z)] 
    + 2\eta k \cos[k(\eta - z)] 
    - 2\sin[k(\eta - z)]
    }{k^3}
    \\
    &\hspace{4cm}
    -
    \frac{
    \eta_{\rm in}^2 k^2 \sin[k(\eta_{\rm in} - z)] 
    + 2\eta_{\rm in} k \cos[k(\eta_{\rm in} - z)] 
    - 2\sin[k(\eta_{\rm in} - z)]
    }{k^3}
    \Bigg\}.
    \nonumber
    \label{eq:amp_ratio_exact}
\end{align}

In the deep sub-horizon limit, 
for a Bunch–Davies initialization in the remote past ($\eta_{\rm in}\to -\infty$) the oscillatory
boundary term averages out, leaving the leading behavior
\begin{equation}
    \ln\!\frac{A_+}{A_-}
    \simeq 
    -\,\frac{2\,\vartheta_0\,q}{l^3}\,\eta^2
    \sin{[k(\eta - z)]},
    \label{eq:amp_ratio_subhorizon}
\end{equation}
for $|k\eta|\gg1$
Thus, the helicity amplitude ratio oscillates with the phase of the CS background, and its envelope scales as
$\eta^2$ in conformal time.  
Since $\eta = -e^{-Ht}/H$ in de Sitter space, this corresponds to an exponentially decaying envelope
$\propto e^{-2Ht}$ in cosmic time.  
The periodic sign change of the sine term indicates alternating amplification and attenuation of
opposite helicities as the gravitational wave propagates through successive CS phase regions.

\subsection{Super-horizon regime}

In the super-horizon regime ($|k\eta|\ll1$), the imaginary part of the frequency splitting
$\Delta\omega$ is given by:
\begin{equation}
    \Im\{\Delta\omega\} = \frac{14\vartheta_{0}}{l^3}\cos{[k(\eta - z)]}.
\end{equation}
Integrating over conformal time we get the relation
\begin{equation}\label{eq:amp_ratio_superhorizon}
    \ln\!\frac{A_+}{A_-} = \frac{14\vartheta_{0}}{l^3 k}\left\{\sin{[k(\eta - z)]}-
    \sin{[k(\eta_{\rm in} - z)]}\right\}
\end{equation}

In the limit of $|k\eta|\ll 1$, we can use the Taylor expansion,
\begin{equation}
\sin[k(\eta - z)] = 
\sin(-kz)
+ k\eta\,\cos(-kz)
- \tfrac{1}{2}(k\eta)^2\sin(-kz)
+ \mathcal{O}[(k\eta)^3].
\end{equation}
Eq.~\eqref{eq:amp_ratio_superhorizon} becomes
\begin{align}
    \ln\!\frac{A_+}{A_-}
    &=
    \frac{14\vartheta_{0}}{l^3 k}
    \left\{\sin(-kz) - \sin[k(\eta_{\rm in} - z)]\right\}
    + \frac{14\vartheta_{0}}{l^3}\,\eta\,\cos(-kz)
    - \frac{7\vartheta_{0}}{l^3}\,k\eta^2\sin(-kz)
    + \mathcal{O}[(k\eta)^2].
    \label{eq:amp_ratio_super_expanded}
\end{align}
The leading term (in French brackets) is independent of $\eta$ and represents a frozen amplitude
asymmetry established when the mode exits the horizon.  
The next term, linear in $\eta$, describes a small residual time dependence suppressed by $k\eta$ and
therefore negligible deep in the super-horizon regime.

The physical interpretation is that
On super-horizon scales, the amplitude birefringence effectively \emph{freezes out};
the differential amplification of the two helicities reaches a constant value once $|k\eta|\ll1$.  
This behaviour reflects the fact that outside the horizon, tensor modes evolve as nearly constant
background distortions, and further parity-violating amplification is exponentially suppressed by
cosmic expansion.

\begin{figure}[h]
    \centering
    \includegraphics[width=0.75\linewidth]{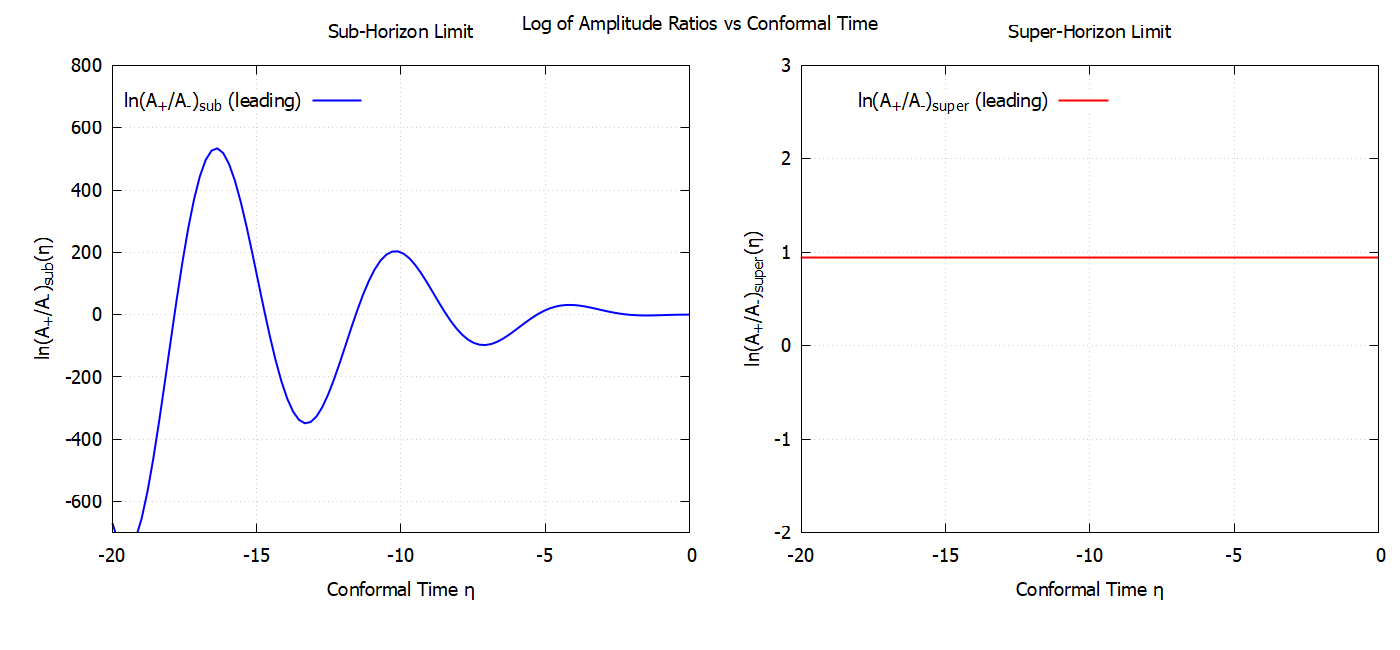}
    \caption{The ratio of the log of amplitudes in the leading order, $\ln\left(A_+/A_-\right)$ versus
    conformal time($\eta$) for both the sub- and super-horizon limits.
    The plot on left for the sub-horizon limit shows the quadratic dependence ($\propto \eta^2 $) and enveloped
    by an oscillatory term. The plot on right for the super-horizon limit shows the constant frozen-in amplitude
     difference that remains etched in the distant future($\eta \rightarrow 0^{-}$)}
    \label{fig:amplitude_birefringence}
\end{figure}

Figure~\ref{fig:amplitude_birefringence} illustrates the logarithm of the helicity amplitude ratio
$\ln(A_+/A_-)$ as a function of conformal time $\eta$ for both the sub-horizon and super-horizon limits.
In the sub-horizon regime, the amplitude ratio exhibits an oscillatory behavior whose envelope grows as
$\eta^2$, reflecting the alternating amplification and attenuation of opposite helicities as they propagate
through the CS background. 
In contrast, the super-horizon behavior approaches a nearly constant value with only a small $\eta$-linear
correction, corresponding to the freezing of amplitude birefringence once the mode exits the horizon. 
The constant offset and the weak slope arise from the residual imaginary part of the frequency splitting,
consistent with the analytic expression derived in eq.~\eqref{eq:amp_ratio_super_expanded}. 
This figure thus highlights the qualitative difference between the dynamic sub-horizon amplification
and the frozen super-horizon asymmetry of gravitational-wave amplitudes in a parity-violating background.

\section{Dark Matter Relations and Modifications}
\label{sec:DM}

The preceding analysis has treated the CS field $\vartheta$ as a massless scalar. However, from a
phenomenological standpoint, it is compelling to consider that $\vartheta$ may itself constitute
the dark matter, or a significant portion of it. Well-motivated candidates for dark matter, such as
axions or axion-like particles (ALPs), are light pseudoscalar fields, and the curvature coupling for
$\vartheta$ has the same discrete symmetries as an ALP coupling.
In this section, we promote $\vartheta$ to a massive field and analyze how its identity as a dark
matter component modifies the gravitational birefringence effects calculated in the massless case.

Building on this idea, it was demonstrated that generalized gravity theories, including scalar-tensor and
string-inspired models, yield distinctive gravitational-wave spectra (typically with a blue tilt, $n_T \simeq 3$)
during pole-like inflation, differing fundamentally from the scale-invariant case in Einstein
gravity~\cite{ref-hwang1998}. It was also shown that the string-theoretic axion coupling to
$^{\star}\mathcal{RR}$ induces parity-violating, polarization-dependent evolution of tensor perturbations,
leading to potentially observable gravitational-wave birefringence imprinted in the CMB~\cite{ref-choi2000}.

\subsection{Massive Chern-Simons dark matter field}

We now consider a massive scalar field with a potential:
\begin{equation}
V(\vartheta) = \frac{1}{2}m^2 \vartheta^2,
\end{equation}
where $m$ is the mass of the dark matter particle. The action from eq.~(\ref{eq-actionS})
remains unchanged, but the equation of motion for the $\vartheta$ field (\ref{eq-theta-motion})
gains a mass term:
\begin{equation}
\Box\vartheta - m^2\vartheta = -\frac{1}{4}(^{\star}\mathcal{RR}).
\label{eq:massive_eom}
\end{equation}
For the homogeneous background field in a de Sitter universe, and neglecting the Pontryagin density source
term at zeroth order, this equation becomes
\begin{equation}
\vartheta'' + 2\frac{\alpha'}{\alpha}\vartheta' + m^2 \alpha^2 \vartheta = 0,
\label{eq:background_massive}
\end{equation}
which is another parametric oscillator equation.
Substituting the scale factor $\alpha = l/\eta$ yields
\begin{equation}
\vartheta'' - \frac{2}{\eta}\vartheta' + \frac{m^2 l^2}{\eta^2} \vartheta = 0.
\end{equation}
The general solution to this equation is a linear combination of Hankel functions,
\begin{equation}
\vartheta(\eta) = (-\eta)^{3/2} \left[ C_1 H^{(1)}_{\nu}(-k\eta) + C_2 H^{(2)}_{\nu}(-k\eta) \right],
\label{eq:gen_sol_massive}
\end{equation}
where
\begin{equation}
\nu = \sqrt{\frac{9}{4} - (m l)^2}.
\end{equation}
The Bunch-Davies vacuum condition selects $C_2 = 0$. The key difference from the massless case is the
order of the Hankel function, $\nu$, which now depends on the product $ml$.

\subsection{Modifications in the sub-horizon limit}

The mass term introduces a new scale that competes with the Hubble parameter. The behavior of the field,
and consequently all derived quantities, changes significantly.

In the sub-horizon limit, the asymptotic form of the solution is a modulated plane wave. However,
the functional form of the amplitude is more complicated than the simple $\eta/l$ decay of the massless
case. The solution can be approximated as:
\begin{equation}
\vartheta_k(\eta) \approx \vartheta_0\mathcal{F}(m l, k \eta) e^{-i k \eta},
\label{eq:approx_sol_sub}
\end{equation}
where $\mathcal{F}(m l, k \eta)$ is a function that reduces to $\eta/l$ for $m=0$ but deviates from
this simple form for $m > 0$.
This altered time dependence propagates into the coefficients that define the Cotton tensor
and the source term. The derivatives $v_\mu = \nabla_\mu \vartheta$ become
\begin{eqnarray}
v_\eta & = & \partial_\eta \vartheta \approx \vartheta_0 (\partial_\eta \mathcal{F} 
- i k \mathcal{F}) e^{-i k \eta} \\
v_z & = & \partial_z \vartheta \approx \vartheta_0 (i k \mathcal{F}) e^{-i k \eta}.
\end{eqnarray}
The second derivatives ($v_{\eta\eta}$, $v_{\eta z}$, and $v_{zz}$) are modified similarly. This means the
Cotton tensor prefactor $f(\omega, q, k, \eta)$ from (\ref{eq-pref-f}) is no longer a simple polynomial in
$\eta$ but a more complex function
\begin{equation}
f(\omega, q, k, \eta) \rightarrow \frac{\vartheta_0}{l}\widetilde{f}(m l, \omega, q, k, \eta).
\end{equation}

The dispersion relation correction in the sub-horizon regime is consequently modified. The frequency
splitting between helicities will become
\begin{equation}
\Delta\omega = \omega_+ - \omega_- \approx \frac{2}{\alpha^2} \left| \widetilde{f}(m l, q, q, k, \eta) \right|.
\label{eq:delta_omega_dm}
\end{equation}
The explicit time dependence $\Delta\omega \propto \eta^2$ is replaced by a mass-dependent function
$\widetilde{f}$. The strength of the birefringence effect in the early universe now explicitly depends
on the dark matter mass $m$. For a very light field ($m \ll H$), we recover the massless result.
For a heavier field, the effect can be either suppressed or enhanced, providing a potential observational
link between the parity-violating imprint on primordial gravitational waves and the mass of the
dark matter particle.

The particular solution for the sourced gravitational waves is also altered. The source term $S(\eta,z)$
scales as $(\nabla\vartheta)^2$. Its new functional form is
\begin{equation}
S(\eta, z) \propto \frac{\vartheta_0^2}{l^2} \, |\mathcal{F}(m l, k \eta)|^2.
\end{equation}
Following the same analysis the amplitude of the particular solution becomes
\begin{equation}
h_{\pm}^{\rm(part)} \propto \pm  \frac{\pi G \vartheta_0 l}{k^2 \eta}|\mathcal{F}(m l, k \eta)|^2.
\label{eq:hsourced_dm}
\end{equation}
This result indicates that the efficiency of gravitational wave production by the CS dark matter field
is mass dependent. A field with some specific mass could potentially be a more efficient generator of
primordial gravitational waves than a massless one.

\subsection{Late-time behavior and consistency}

A crucial feature of this model is its consistency with the standard cosmological model at
late times. In the frozen regime ($H\gg m)$, $\vartheta$ is nearly
constant and maximally sources parity-violating birefringence.
When the Hubble parameter drops below the mass scale ($H < m$), the field $\vartheta$
begins to oscillate, and its energy density redshifts as pressureless dust, $\rho_{\vartheta} \propto a^{-3}$,
behaving as standard cold dark matter~\cite{ref-turner,ref-sikivie,ref-marsh}.
Furthermore, the amplitude of oscillation decays as $\langle \vartheta \rangle \propto a^{-3/2}$.
Since the CS effects are driven by derivatives of $\vartheta$, they become exponentially
suppressed at late times. This ensures that the intense parity-violating effects are confined
to the early universe, evading constraints from solar system tests, binary pulsar observations,
and direct gravitational wave detections by LIGO/Virgo/KAGRA~\cite{ref-yunes2009,ref-alexander2009}.

The model leaves behind a permanent, frozen record of this early-era physics in the form of
a chiral, blue-tilted primordial gravitational wave spectrum, without affecting late-time
gravity. This connects naturally with studies of parity violation in the early universe and its
possible observational imprints on the CMB and stochastic
gravitational wave background~\cite{ref-kamionkowski}. In this way,
the birefringence scaling can act as a probe of the dark matter mass, interpolating smoothly
between the massless axion-like case and the standard cold dark matter regime.

Promoting the CS field $\vartheta$ to a massive dark matter candidate introduces a potentially
rich layer of phenomenology. The simple, power-law time dependencies found in the massless
case are replaced by more complex, mass-dependent functions. This establishes a direct
theoretical link between the properties of dark matter (its mass $m$) and the characteristics
of primordial gravitational birefringence and wave generation. This model could provide a
compelling framework for seeking dark matter signatures in the polarization of the cosmic
microwave background and the stochastic gravitational wave background, and this connection will be
examined in future work.

\section{Conclusion}
\label{sec:Conclusion}

In this work, we have carried out a detailed analysis of tensor perturbations in de Sitter spacetime within the framework of CS modified gravity. Beginning from the conformally flat form of the metric, we computed the curvature tensors and perturbed field equations, showing explicitly that the Pontryagin density vanishes at linear order while the Cotton tensor introduces parity-violating corrections to gravitational wave propagation. The decoupled helicity modes reveal birefringence in both sub- and super-horizon regimes: oscillatory and exponentially suppressed inside the horizon, but frozen to a constant splitting once the modes cross the horizon. We also derived particular solutions sourced by the scalar background, demonstrating the amplification of one helicity due to the CS coupling, together with explicit expressions for the radiated flux and its smooth reduction to the flat-space limit.

A key observable studied here is the accumulated phase difference between right- and left-handed tensor modes. We showed that in the sub-horizon limit, this phase difference grows quadratically with conformal time, whereas in the super-horizon regime it approaches a frozen value that survives as a persistent parity-violating imprint in the primordial tensor spectrum. This provides a clear physical mechanism for how CS modifications can be encoded in the polarization of relic gravitational waves, consistent with earlier suggestions that such parity-violating effects could leave observable imprints on the CMB and stochastic backgrounds. Notably, the birefringence effect exhibits a continuous transition from amplitude modulation to velocity (phase) birefringence as the scalar field phase evolves---indicating that the dominant observable shifts from differential damping to differential propagation speed depending on the relative phase between the gravitational and scalar fields.

Finally, we extended the analysis by promoting the CS scalar to a massive dark matter candidate. The introduction of a mass modifies both the time dependence of the background field and the efficiency of gravitational wave sourcing, leading to a direct link between dark matter phenomenology and parity-violating gravitational imprints. Importantly, the CS effects are naturally suppressed at late times, ensuring consistency with astrophysical and gravitational wave constraints, while leaving unique signatures in the early universe.

In summary, our results demonstrate that Chern-Simons modified gravity in a de Sitter background produces distinctive birefringence and phase-shift signatures in primordial tensor modes, potentially accessible to future cosmic microwave background polarization measurements and gravitational wave observatories. Future work should address higher-order effects, nonlinear backreaction, and realistic reheating scenarios to refine these predictions and strengthen the observational connection between parity-violating gravity and early-universe cosmology.

\newpage

\appendix

\section{Appendix: Background Metric Calculations}

For completeness, we provide the detailed calculations of the background de Sitter metric used in the main text.

The conformal form of the de Sitter metric is
\begin{equation}
ds^2 = \alpha(\eta)^2 \left( -d\eta^2 + \delta_{ij}\,dx^{i}dx^{j} \right),
\end{equation}
with $\alpha(\eta)=l/\eta$. The metric elements are thus
\begin{equation}
g_{\mu\nu} = \alpha^2 \,\mathrm{diag}(-1,1,1,1), \quad
g^{\mu\nu} = \alpha^{-2} \,\mathrm{diag}(-1,1,1,1).
\end{equation}

From the definition
\begin{equation}
\Gamma^{\lambda}_{\mu\nu} = \tfrac{1}{2} g^{\lambda\sigma} 
\left( \partial_\nu g_{\mu\sigma} + \partial_\mu g_{\nu\sigma} - \partial_\sigma g_{\mu\nu} \right),
\end{equation}
the non-vanishing Christoffel symbols are
\begin{align}
\Gamma^{\eta}_{\eta\eta} &= \frac{\alpha'}{\alpha} = -\frac{1}{\eta}, \\
\Gamma^{\eta}_{ij} &= \frac{\alpha'}{\alpha} \delta_{ij} = -\frac{1}{\eta}\delta_{ij}, \\
\Gamma^{i}_{\eta j} &= \frac{\alpha'}{\alpha}\delta^i_j = -\frac{1}{\eta}\delta^i_j.
\end{align}

The Riemann tensor components are
\begin{align}
R^\eta_{\ i \eta j} &= \left( \frac{\alpha''}{\alpha} - \frac{\alpha'^2}{\alpha^2} \right)\delta_{ij}, \\
R^i_{\ jkl} &= \left( \frac{\alpha'}{\alpha} \right)^2 \left( \delta^i_k \delta_{jl} -
\delta^i_l \delta_{jk} \right).
\end{align}
From this, the Ricci tensor becomes
\begin{align}
R_{\eta\eta} &= -d\left( \frac{\alpha''}{\alpha} - \frac{\alpha'^2}{\alpha^2} \right) = -\frac{d}{\eta^2}, \\
R_{ij} &= \left( \frac{\alpha''}{\alpha} - \frac{\alpha'^2}{\alpha^2} +
(d-1)\frac{\alpha'^2}{\alpha^2} \right)\delta_{ij} 
= \frac{d}{\eta^2}\delta_{ij}.
\end{align}
The Ricci scalar is
\begin{equation}
R = \frac{d(d+1)}{l^2}.
\end{equation}

Finally, the Einstein tensor in $d+1$ dimensions is
\begin{align}
G_{\eta\eta} &= \frac{d(d-1)}{2\eta^2}, \\
G_{ij} &= -\frac{d(d-1)}{2\eta^2}\delta_{ij}, \\
G_{\eta i} &= 0.
\end{align}

These results confirm the maximally symmetric nature of the de Sitter spacetime and serve as the
background quantities for the perturbative analysis in the main text.

\section{Appendix: Perturbed Metric Calculations}
\label{sec-perturbation}

The perturbed metric of the de Sitter space is defined as
\begin{equation}
    ds^2 = \alpha(\eta)^2\left[-d\eta^2 + (\delta_{ij} + h^{TT}_{ij})dx^i\,dx^j\right],
\end{equation}
where $h^{TT}_{ij}$ is defined as the traceless-transverse (TT) tensor modes describing gravitational waves.
The particular TT-tensor mode used in this paper is
\begin{equation}
    h^{TT}_{ij} = \begin{pmatrix}
        h_{+} & h_{\times} & 0\\
        h_{\times} & -h_{+} & 0\\
        0 & 0 & 0\\
    \end{pmatrix},
\end{equation}
which is a wave propagating in the $z$-direction
[i.e. $h_{+} = h_{+}(\eta - z)$ and $h_{\times} = h_{\times}(\eta - z)$].
Using this we calculate the Christoffel symbols. The non-zero Christoffel symbols are
\begin{eqnarray}
        \Gamma^{\eta}_{\eta\eta} & = & \frac{1}{2}g^{\eta\eta}(\partial_{\eta}g_{\eta\eta})
        =\frac{\alpha '}{\alpha}\\
        \Gamma^{\eta}_{ij} & = & \frac{1}{2}g^{\eta\eta}(-\partial_{\eta}g_{ij})
        = \frac{\alpha '}{\alpha}\delta_{ij} + \frac{1}{2\alpha^2}\partial_{\eta}(\alpha^2\gamma_{ij})\\
        \Gamma^{i}_{\eta j} & = & \frac{1}{2}g^{im}(\partial_{\eta}g_{mj})
        = \frac{\alpha '}{\alpha}\delta^{i}_{j} + \frac{1}{2\alpha^2}\partial_{\eta}(\alpha^2\gamma^{i}_{j})\\
        \Gamma^{i}_{jk} & = & \frac{1}{2}g^{im}\left(g_{jm,k} + g_{mk,j} - g_{jk,m}\right)
        = \frac{1}{2}\left(\partial_{k}\gamma^{i}_{j} + \partial_{j}\gamma^{i}_{k}
        - \partial^{i}\gamma_{jk}\right),
\end{eqnarray}
where the spatial metric $\gamma_{ij} = h^{TT}_{ij}$, and 
we have used the fact that after perturbation the spatial metric elements take the form
\begin{equation}
    g_{ij} = \alpha^2(\delta_{ij} + \gamma_{ij})\quad\implies\quad g^{ij} = \alpha^{-2}\delta^{ij}
\end{equation}
In linearized gravity we can further simplify the Christoffel symbols as follows:
\begin{eqnarray}
        \Gamma^{\eta}_{ij} & = & \frac{\alpha '}{\alpha}\delta_{ij}
        + \frac{1}{2\alpha^2}\partial_{\eta}(\alpha^2\gamma_{ij})\\
        & = & \frac{\alpha '}{\alpha}\delta_{ij} + \frac{\alpha '}{\alpha}\gamma_{ij}
        + \frac{1}{2}\partial_{\eta}\gamma_{ij}.
\end{eqnarray}
In linearized gravity, we keep only the first order terms in perturbations. ($\gamma_{ij}$ is of the first
order.) The term $\frac{\alpha '}{\alpha}\gamma_{ij}$ is second order and is typically dropped since
it is a product of the background term and the first order perturbation leading to nonlinear back-reaction
effects. So, up to first order in the perturbation, the Christoffel symbol may be written as
\begin{equation}
    \Gamma^{\eta}_{ij} \approx \frac{\alpha '}{\alpha}\delta_{ij} +  \frac{1}{2}\partial_{\eta}\gamma_{ij}
    + \mathcal{O}(\gamma^2).
\end{equation}
The Christoffel symbols up to the first order in the perturbation are
\begin{eqnarray}
            \Gamma^{\eta}_{\eta\eta} & = & \frac{\alpha '}{\alpha} = \frac{-1}{\eta}\\
            \Gamma^{\eta}_{ij} & = & \frac{\alpha '}{\alpha}\delta_{ij} +  \frac{1}{2}\partial_{\eta}\gamma_{ij}
            = \frac{-1}{\eta}\delta_{ij} + \frac{1}{2}\partial_{\eta}\gamma_{ij}\\
            \Gamma^{i}_{\eta j} & = & \frac{\alpha '}{\alpha}\delta^{i}_{j} +
            \frac{1}{2}\partial_{\eta}\gamma^{i}_{j} = \frac{-1}{\eta}\delta^{i}_{j} 
            \frac{1}{2}\partial_{\eta}\gamma^{i}_{j}\\
            \Gamma^{i}_{jk} & = & \frac{1}{2}\left(\partial_{k}\gamma^{i}_{j} + \partial_{j}\gamma^{i}_{k}
            - \partial^{i}\gamma_{jk}\right).
\end{eqnarray}

Using the forma of the linearized Christoffel symbols, we can calculate the general Riemann tensor terms.
For a TT gauge, the take the forms
\begin{eqnarray}
        \tensor{R}{^{\eta}_{i\eta j}} & = & \partial_{\eta}\Gamma^{\eta}_{ij} +
        \Gamma^{\eta}_{\eta\lambda}\Gamma^{\lambda}_{ij} - \Gamma^{\eta}_{\lambda j}\Gamma^{\lambda}_{i\eta}\\
        & = & \left[\frac{\alpha ''}{\alpha} - \frac{(\alpha ')^2}{\alpha^2}\right]\delta_{ij} +
        \frac{1}{2}\partial^2_{\eta}\gamma_{ij} + \frac{\alpha '}{2\alpha}\partial_{\eta}\gamma_{ij} +
        \mathcal{O}(\gamma^2)\\
        \tensor{R}{^{\eta}_{ijk}} & = & \partial_{j}\Gamma^{\eta}_{ik} - \partial_{k}\Gamma^{\eta}_{ij}
        + \Gamma^{\eta}_{j\lambda}\Gamma^{\lambda}_{ik} - \Gamma^{\eta}_{k\lambda}\Gamma^{\lambda}_{ij}\\
        & = & \frac{1}{2}\left(\partial_{j}\partial_{\eta}\gamma_{ik} -
        \partial_{k}\partial_{\eta}\gamma_{ij}\right) +  \mathcal{O}(\gamma^2)\\
        \tensor{R}{^{i}_{j\eta k}} & = & \partial_{\eta}\Gamma^{i}_{jk} - \partial_{k}\Gamma^{i}_{j\eta}
        + \Gamma^{i}_{\eta\lambda}\Gamma^{\lambda}_{jk} - \Gamma^{i}_{k\lambda}\Gamma^{\lambda}_{j\eta}\\
        & = & \frac{1}{2}\partial_{\eta}\left(\partial_{j}\gamma^{i}_{k} - \partial^{i}\gamma_{jk}\right)
        - \frac{\alpha'}{\alpha}\partial_{k}\gamma^{i}_{j} + \mathcal{O}(\gamma^2)\\
        \tensor{R}{^{i}_{jkl}} & = & \partial_{k}\Gamma^{i}_{jl} - \partial_{l}\Gamma^{i}_{jk} +
        \Gamma^{i}_{k\lambda}\Gamma^{\lambda}_{jl} - \Gamma^{i}_{l\lambda}\Gamma^{\lambda}_{jk}\\
        & = & \frac{1}{2}\left(\partial_{k}\partial_{j}\gamma^{i}_{l} - \partial_{k}\partial^{i}\gamma_{jl}
        - \partial_{l}\partial_{j}\gamma^{i}_{k} + \partial_{l}\partial^{i}\gamma_{jk}\right)
        + \left(\frac{\alpha '}{\alpha}\right)^2\left(\delta^{i}_{k}\delta_{jl} -
        \delta^{i}_{l}\delta_{jk}\right) \\
        & & + \frac{\alpha '}{2\alpha}\left(\delta^{i}_{k}\partial_{\eta}\gamma_{jl} +
        \delta_{jl}\partial_{\eta}\gamma^{i}_{k} - \delta^{i}_{l}\partial_{\eta}\gamma_{jk} -
        \delta_{jk}\partial_{\eta}\gamma^{i}_{l}\right) + \mathcal{O}(\gamma^2). \nonumber
\end{eqnarray}
We know that the perturbations are only functions of $\eta$ and $z$, the propagation direction,
and using the fact that the perturbation is traceless and transverse, we can simplify the nonvanishing Riemann
terms to get:
\begin{eqnarray}
        \tensor{R}{^{\eta}_{i\eta j}} & = & \left[\frac{\alpha ''}{\alpha} -
        \frac{(\alpha ')^2}{\alpha^2}\right]\delta_{ij} + \frac{1}{2}\partial^2_{\eta}\gamma_{ij}
        + \frac{\alpha '}{2\alpha}\partial_{\eta}\gamma_{ij}\\
        \tensor{R}{^{\eta}_{izj}} & = & \frac{1}{2}\partial_{z}\partial_{\eta}\gamma_{ij}\\
        \tensor{R}{^{i}_{z\eta j}} & = & \frac{1}{2}\partial_{\eta}\partial_{z}\gamma^{i}_{j}\\
        \tensor{R}{^{i}_{j\eta z}} & = & -\frac{\alpha'}{\alpha}\partial_{z}\gamma^{i}_{j}\\
        \tensor{R}{^{i}_{zzj}} & = & \frac{1}{2}\partial^2_{z}\gamma^{i}_{j}
        -\left(\frac{\alpha'}{\alpha}\right)^2\delta^{i}_{j}
        -\frac{\alpha'}{2\alpha}\partial_{\eta}\gamma^{i}_{j}.
\end{eqnarray}

The Ricci tensor may then be calculated, starting with the time-time component,
\begin{eqnarray}
        R_{\eta\eta} & = & \tensor{R}{^{\lambda}_{\eta\lambda\eta}} =
        \cancelto{0}{\tensor{R}{^{\eta}_{\eta\eta\eta}}} + \tensor{R}{^{i}_{\eta i\eta}}
        =-\tensor{R}{^{\eta}_{i\eta i}}\\
        & = & -d\left\{\left[\frac{\alpha ''}{\alpha} - \frac{(\alpha ')^2}{\alpha^2}\right] +
        \frac{1}{2}\partial^2_{\eta}\cancelto{0}{\gamma_{ii}}\quad
        - \frac{\alpha '}{2\alpha}\partial_{\eta}\cancelto{0}{\gamma_{ii}}\quad\right\},\\
        & = & -\frac{d}{\eta^2} = \bar{R}_{\eta\eta}.
\end{eqnarray}
where we have taken advantage of the TT gauge and left the spatial dimension $d\,(=3)$ explicit.    
Similarly, for the spatial components of the Ricci tensor, we find
\begin{eqnarray}
        R_{ij} & = & \tensor{R}{^{\lambda}_{i\lambda j}} = \tensor{R}{^{\eta}_{i\eta j}}
        +\tensor{R}{^{k}_{ikj}}\\
        & = & \left[\frac{\alpha ''}{\alpha} - \frac{(\alpha ')^2}{\alpha^2}\right]\delta_{ij}
        + \frac{1}{2}\partial^2_{\eta}\gamma_{ij} 
        + \frac{\alpha '}{2\alpha}\partial_{\eta}\gamma_{ij} \nonumber\\
        & & +\left(\frac{\alpha'}{\alpha}\right)^2(d - 1)\delta_{ij}
        -\frac{1}{2}\partial_{k}\partial^{k}\gamma_{ij}
        + \frac{\alpha'}{2\alpha}(d  -2)\partial_{\eta}\gamma_{ij}\\
        & = & \frac{d}{\eta^2}\delta_{ij} + \frac{1}{2}\left[\partial^2_{\eta}\gamma_{ij}
        + (d  -1)\frac{\alpha'}{\alpha}\partial_{\eta}\gamma_{ij}
        - \partial_{k}\partial^{k}\gamma_{ij}\right] \\
        & = & \bar{R}_{ij} + \frac{1}{2}\left[\partial^2_{\eta}\gamma_{ij} +
        (d  -1)\frac{\alpha'}{\alpha}\partial_{\eta}\gamma_{ij}  - \partial_{k}\partial^{k}\gamma_{ij}\right].
\end{eqnarray}
The time-space components of the Ricci tensor are easily seen to vanish because of the TT condition,
\begin{equation}
    R_{\eta i} = R_{i \eta} = 0.
\end{equation}

The Ricci scalar in $d + 1$ dimensions is, up to first order in the perturbations,
\begin{equation}
        \mathcal{R} = g^{\mu\nu}R_{\mu\nu} = g^{\eta\eta}R_{\eta\eta} + g^{ij}R_{ij}
        = -\alpha^{-2}\,R_{\eta\eta} + \alpha^{-2}\delta^{ij}\,R_{ij}.
\end{equation}
Using the explicit forms for the diagonal components to the Ricci tensor to take the trace,
\begin{eqnarray}
        \mathcal{R} & = & \frac{d}{\alpha^2\eta^2} + \frac{1}{\alpha^2}\delta^{ij}
        \left\{\frac{d}{\eta^2}\delta_{ij} + \frac{1}{2}\left[\partial^2_{\eta}\gamma_{ij} +
        (d - 1)\frac{\alpha'}{\alpha}\partial_{\eta}\gamma_{ij} -  \nabla^2\gamma_{ij}\right]\right\}\\
        & = & \frac{d}{\eta^2\alpha^2} + \frac{d^2}{\alpha^2\eta^2} +
        \delta^{ij}\left[\partial^2_{\eta}\gamma_{ij} + (d - 1)\frac{\alpha'}{\alpha}\partial_{\eta}\gamma_{ij}
        -  \nabla^2\gamma_{ij}\right].
\end{eqnarray}
The TT condition gives us that $\delta^{ij}\gamma_{ij} = 0$. Using this, the second term in the relation
above becomes
\begin{equation}
    \alpha^{-2}\delta^{ij}\partial^2_{\eta}\gamma_{ij} = \partial^2_{\eta}(\delta^{ij}\gamma_{ij}) = 0.
\end{equation}
Similarly, the other terms with $\gamma_{ij}$ also vanish. So the Ricci scalar takes the form
\begin{equation}
        \mathcal{R} = \frac{d(d +1)}{\eta^2\alpha^2} = \frac{d(d + 1)}{l^2}.
\end{equation}

With the Ricci tensor and scalar in hand, the Einstein tensor components are straightforward to calculate.
The time-time components is
\begin{eqnarray}
        G_{\eta\eta} & = & R_{\eta\eta} - \frac{1}{2}\mathcal{R}g_{\eta\eta}\\
        & = & -\frac{d}{\eta^2} - \frac{1}{2} \frac{d(d + 1)}{\eta^2\alpha^2}(-\alpha^2)\\
        & = & \frac{d(d  -1)}{2\eta^2} = \bar{G}_{\eta\eta},
\end{eqnarray}
and the space-space,
\begin{eqnarray}
        G_{ij} & = &  R_{ij} - \frac{1}{2}\mathcal{R}g_{ij}\\
        & = & \left(\frac{d}{\eta^2}\right)\delta_{ij} + \frac{1}{2}\left[\partial^2_{\eta}\gamma_{ij}
        +(d - 1)\frac{\alpha'}{\alpha}\partial_{\eta}\gamma_{ij} - \nabla^2\gamma_{ij}\right]
        - \frac{1}{2} \frac{d(d + 1)}{\eta^2\alpha^2}[\alpha^2(\delta_{ij} + \gamma_{ij})]\\
        & = & \Bar{G}_{ij} +  \frac{1}{2}\left[\partial^2_{\eta}\gamma_{ij}
        + (d - 1)\frac{\alpha'}{\alpha}\partial_{\eta}\gamma_{ij} - \nabla^2\gamma_{ij}\right],   
\end{eqnarray}
while, obviously, the time-space components vanish,
\begin{equation}
G_{\eta i} = G_{i\eta} = 0.
\end{equation}

\section{Appendix: Cotton Tensor Calculations}

In a CS background we get an additional second-rank tensor in the field equations, now commonly known as
the Cotton tensor (representing a generalization of the original three-dimensional Cotton tensor). The
Cotton tensor has two parts associated with it---one which is associated with the symmetries of the Ricci
tensor and the other which contains the dual of the Riemann tensor. We have denoted them as the first and
second parts of the Cotton tensor below. In terms of Riemann tensor components, we have
\begin{eqnarray}
        C^{\mu\nu}_{(1)} & = & -\frac{1}{2}v_{\alpha}\left(\epsilon^{\alpha\mu\sigma\tau}\nabla_{\sigma}
        R^{\nu}_{\tau} + \epsilon^{\alpha\nu\sigma\tau}\nabla_{\sigma}R^{\mu}_{\tau}\right)\\
        C^{\mu\nu}_{(2)} & = & -\frac{1}{2}v_{\sigma\tau}\left(^{\star}R^{\tau\mu\sigma\nu}
        +^{\star}R^{\tau\nu\sigma\mu}\right),
\end{eqnarray}
where we have used $v_{\sigma\tau} = \nabla_{\sigma}v_{\tau}$. Calculations of the first order perturbation
forms of these terms are shown below, beginning with the first part of the Cotton tensor,
\begin{eqnarray}
        C^{\mu\nu}_{(1)} & = & -\frac{1}{2}v_{\alpha}\left(\epsilon^{\alpha\mu\sigma\tau}\nabla_{\sigma}
        R^{\nu}_{\tau} + \epsilon^{\alpha\nu\sigma\tau}\nabla_{\sigma}R^{\mu}_{\tau}\right)\\
        & = & -\frac{1}{2}v_{\alpha}\left[\epsilon^{\alpha\mu\sigma\tau}\nabla_{\sigma}\left(R^{\nu}_{\tau}
        + \delta R^{\nu}_{\tau}\right) + \epsilon^{\alpha\nu\sigma\tau}\nabla_{\sigma}
        \left(R^{\mu}_{\tau} + \delta R^{\mu}_{\tau}\right)\right]\\
        & = & \Bar{C}^{\mu\nu}_{(1)} - \frac{\epsilon}{2}\left[v_{\alpha}
        \left(\epsilon^{\alpha\mu\sigma\tau}\nabla_{\sigma}\delta R^{\nu}_{\tau}
        +\epsilon^{\alpha\nu\sigma\tau}\nabla_{\sigma}\delta R^{\mu}_{\tau}\right)\right].
\end{eqnarray}
This expresses the first part of the Cotton tensor as
$C^{\mu\nu}_{(1)} = \Bar{C}^{\mu\nu}_{(1)} + \epsilon\delta C^{\mu\nu}_{(1)}$,
where the perturbation term $\delta C^{\mu\nu}_{(1)}$ can be also written as
\begin{equation}
    \delta C^{(1)}_{\mu\nu} = -\frac{1}{2}\left[v_{\alpha}\left(\epsilon^{\alpha\beta\sigma\tau}
    \Bar{g}_{\mu\beta}\nabla_{\sigma}\delta R_{\nu\tau} + \epsilon^{\alpha\gamma\sigma\tau}
    \Bar{g}_{\nu\gamma}\nabla_{\sigma}\delta R_{\mu\tau}\right)\right].
\end{equation}
Similarly, for the second part of Cotton tensor we get
\begin{eqnarray}
        C^{\mu\nu}_{(2)} & = & -\frac{1}{2}v_{\sigma\tau}\left(^{\star}R^{\tau\mu\sigma\nu}
        + ^{\star}R^{\tau\nu\sigma\mu}\right)\\
        & = & -\frac{1}{2}v_{\sigma\tau}\left(g^{\mu\lambda}\,^{\star}\tensor{R}{^\tau_\lambda^{\sigma\nu}}
        +g^{\nu\lambda}\, ^{\star}\tensor{R}{^\tau_\lambda^{\sigma\mu}}\right)\\
        & = & -\frac{1}{2}v_{\sigma\tau}\left(\frac{1}{2}g^{\mu\lambda}\epsilon^{\sigma\nu\alpha\beta}
        \tensor{R}{^\tau _{\lambda\alpha\beta}} + \frac{1}{2}g^{\nu\lambda}\epsilon^{\sigma\mu\alpha\beta}
        \tensor{R}{^\tau _{\lambda\alpha\beta}}\right)\\
        & = & -\frac{1}{4}v_{\sigma\tau}\tensor{R}{^\tau _{\lambda\alpha\beta}}
        \left[\left(\Bar{g}^{\mu\lambda}\epsilon^{\sigma\nu\alpha\beta}
        + \Bar{g}^{\nu\lambda}\epsilon^{\sigma\mu\alpha\beta} \right)
        + \epsilon\left(h^{\mu\lambda}\epsilon^{\sigma\nu\alpha\beta}
        + h^{\nu\lambda}\epsilon^{\sigma\mu\alpha\beta}\right)\right].
\end{eqnarray}
Expand the Riemann tensor terms as $\tensor{R}{^\tau _{\lambda\alpha\beta}}
=\tensor{\Bar{R}}{^\tau _{\lambda\alpha\beta}}+\epsilon\delta\tensor{R}{^\tau _{\lambda\alpha\beta}}$,
we get
\begin{equation}
    C^{\mu\nu}_{(2)} = \Bar{C}^{\mu\nu}_{(2)}
    -\frac{\epsilon}{4}\left[v_{\sigma\tau}\delta\tensor{R}{^\tau _{\lambda\alpha\beta}}
    \left(\Bar{g}^{\mu\lambda}\epsilon^{\sigma\nu\alpha\beta}
    +\Bar{g}^{\nu\lambda}\epsilon^{\sigma\mu\alpha\beta} \right)\right] + \mathcal{O}(\epsilon^2).
\end{equation}
When we write the Cotton tensor as
$C_{\mu\nu} = \Bar{C}_{\mu\nu} + \epsilon \delta C_{\mu\nu} + \mathcal{O}(\epsilon^2)$,
this can be also written as
\begin{equation}
    \delta C^{(2)}_{\mu\nu} = -\frac{1}{4}v_{\sigma\tau}
    \left[\Bar{g}_{\nu\gamma}\epsilon^{\sigma\gamma\alpha\beta}\delta\tensor{R}{^\tau _{\mu\alpha\beta}}
    +\Bar{g}_{\mu\rho}\epsilon^{\sigma\rho\alpha\beta}\delta\tensor{R}{^\tau _{\nu\alpha\beta}}\right],
\end{equation}
making the full expression for the first-order perturbation of the Cotton tensor
\begin{eqnarray}
    \delta C_{\mu\nu} & = & -\left[\frac{1}{2}v_{\alpha}\left(\epsilon^{\alpha\beta\sigma\tau}
    \Bar{g}_{\mu\beta}\nabla_{\sigma}\delta R_{\nu\tau} + \epsilon^{\alpha\gamma\sigma\tau}
    \Bar{g}_{\nu\gamma}\nabla_{\sigma}\delta R_{\mu\tau}\right)\right. \nonumber\\
    & & + \left.\frac{1}{4}v_{\sigma\tau}\left(\Bar{g}_{\nu\gamma}\epsilon^{\sigma\gamma\alpha\beta}
    \delta\tensor{R}{^\tau _{\mu\alpha\beta}} + \Bar{g}_{\mu\rho}\epsilon^{\sigma\rho\alpha\beta}
    \delta\tensor{R}{^\tau _{\nu\alpha\beta}}\right)\right].
    \label{eq:Cotton_pert}
\end{eqnarray}
From this general expression, we may derive the forms of specific tensor components.

\subsection{Time-time Cotton tensor}

Using the fact that $\bar{C}_{\eta\eta} = 0$, we can find the perturbation in the time-time Cotton tensor
comnponent,
\begin{equation}
    \delta C_{\eta\eta} = -\left[v_{\alpha}\left(\epsilon^{\alpha\beta\sigma\tau}
    \Bar{g}_{\eta\beta}\nabla_{\sigma}\delta R_{\eta\tau}\right)
    +\frac{1}{2}v_{\sigma\tau}\left(\Bar{g}_{\eta\gamma}\epsilon^{\sigma\gamma\alpha\beta}
    \delta\tensor{R}{^\tau _{\eta\alpha\beta}} \right)\right].
\end{equation}
For the first term on the right-hand side, we can see that the term will exist only when $\beta = \eta$
and also $\tau = \eta$; however, if both of these indices are taken to be $\eta$ then the Levi-Civita
symbol vanishes. So the first does not contribute. For the second term on the right-hand side, we see that
$\gamma = \eta$, which makes all the other indices $\sigma, \alpha$, and $\beta$ spatial. Therefore,
the expression becomes
\begin{equation}
    \delta C_{\eta\eta} = -\frac{1}{2}v_{m\tau}\bar{g}_{\eta\gamma}\epsilon^{m\gamma ij}
    \delta\tensor{R}{^{\tau}_{\eta ij}}.
\end{equation}
The Riemann tensor has the summable index $\tau$ which can take either of the values $\eta$ or a spatial index.
When we substitute $\tau = \eta$ the Riemann tensor $\tensor{R}{^{\eta}_{\eta ij}}$ vanishes. So the only
potentially existent term that remains is
\begin{eqnarray}
     \delta C_{\eta\eta} & = & -\frac{1}{2}v_{mk}\bar{g}_{\eta\gamma}\epsilon^{m\gamma ij}
     \delta\tensor{R}{^{k}_{\eta ij}}\\
     & = & -\frac{1}{4}v_{mk}\bar{g}_{\eta\gamma}\epsilon^{m\gamma ij}
     \left(\partial_{i}\partial_{\eta}\gamma^{k}_{j} - \partial_{j}\partial_{\eta}\gamma^{k}_{i}\right),
\end{eqnarray}
which also vanished because of the TT condition. So the final expression for the time-time Cotton tensor
component is
\begin{equation}
    \delta C_{\eta\eta} = 0.
\end{equation}

\subsection{Mixed Cotton tensor}

For the mixed time-space components in the perturbed metric, we again calculate the two parts, starting with
\begin{equation}
    \delta C^{(1)}_{\eta i} = -\frac{1}{2}\left[v_{\alpha}\left(\epsilon^{\alpha\beta\sigma \tau}
    \Bar{g}_{\eta\beta}\nabla_{\sigma}\delta R_{i\tau} + \epsilon^{\alpha\gamma\sigma \tau}
    \Bar{g}_{i\gamma}\nabla_{\sigma}\delta R_{\eta\tau}\right)\right].
\end{equation}
The second term on the right-hand side demands that $\tau = \eta$, and that also means that $\sigma = \eta$,
since as $R_{\eta\eta}$ is a function of $\eta$ only. That makes the Levi-Civita symbol become zero; hence
the second term makes no contribution. For the first term to exist, the only nonzero contribution will come
if $\tau = j$, leaving the components to take the form
\begin{eqnarray}
        \delta C^{(1)}_{\eta i} & = & -\frac{1}{2}v_{\alpha}\left(\epsilon^{\alpha\beta\sigma j}
        \Bar{g}_{\eta\beta}\nabla_{\sigma}\delta R_{ij}\right)\\
        & = & -\frac{\alpha^2}{2}\,v_{\alpha}\left(\epsilon^{\alpha\beta\sigma j}
        \delta_{\eta\beta}\nabla_{\sigma}\delta R_{ij}\right).
\end{eqnarray}
Since $\beta$ has to be replaced by $\eta$ as demanded by $\delta_{\eta\beta}$,
this implies $\alpha$ and $\sigma$ have to be spatial indices,
and we know that $\alpha$ has to be $z$ for this to exist. Hence, the relation takes the form
\begin{equation}
    \delta C^{(1)}_{\eta i} = \frac{\alpha^2}{2}\,v_{z}
    \left(\delta_{\eta\beta}\epsilon^{z\beta k j}\nabla_{k}\delta R_{ij}\right).
\end{equation}
Again, for this to exist $k$ has to be $z$, and that leads to two repeated indices in the Levi-Civita
symbol, which thus vanishes. Hence, we get that the first part of the perturbation in the mixed Cotton
tensor vanishes.
For the second part of the perturbed Cotton tensor we have the expression
\begin{equation}
    C^{(2)}_{\eta i} = -\frac{1}{4}v_{\sigma\tau}\left[\bar{g}_{i\gamma}
    \epsilon^{\sigma\gamma\alpha\beta}\delta\tensor{R}{^{\tau}_{\eta\alpha\beta}}+
    \bar{g}_{\eta\rho}\epsilon^{\sigma\rho\alpha\beta}\delta\tensor{R}{^{\tau}_{i\alpha\beta}}\right].
\end{equation}
Using the sets of indices for $\sigma$ and $\tau$ that are possible,
we see that this second part also vanishes. Hence, we get another three vanishing tensor components,
\begin{equation}
    \delta C_{\eta i} = 0.
\end{equation}

\subsection{Spatial Cotton tensor}

Things become more complicated with the space-space terms in the Cotton tensor.
The background Cotton tensor for the unperturbed de Sitter spacetime was calculated to vanish.
Therefore, the only contribution will be from the perturbation part of the metric. However, unlike
$\delta C_{\eta\eta}$ and $\delta C_{\eta i}$, we need not have $\delta C_{ij}$ be zero.
We can see in the first term on the right-hand side of eq.~\eqref{eq:Cotton_pert} with $\mu=i$
and $\nu=j$ that the only existent perturbation in the Ricci tensor occurs when we have $\tau$ as another
spatial index. No other Ricci tensor terms have any perturbation. Using this we get
\begin{equation}
    \delta C^{(1)}_{ij} = -\frac{1}{2}\left[v_{\alpha}
    \left(\epsilon^{\alpha\beta\sigma k}\Bar{g}_{i\beta}\nabla_{\sigma}\delta R_{jk}
    + \epsilon^{\alpha\gamma\sigma k}\Bar{g}_{j\gamma}\nabla_{\sigma}\delta R_{ik}\right)\right].
\end{equation}
There can be two possible sets of indices that keep the Levi-Civita symbol intact: $\alpha = \eta$ and
$\sigma = z$, or $\alpha = z$ and $\sigma = \eta$. Combining those two, we get
\begin{equation}
    \delta C^{(1)}_{ij} = -\frac{1}{2}\left(v_{\eta}\nabla_{z} - v_{z}\nabla_{\eta}\right)
    \left(\bar{g}_{i\rho}\epsilon^{\eta\rho zk}\delta R_{jk}
    +\bar{g}_{j\gamma}\epsilon^{\eta\gamma zk}\delta R_{ik}\right).
\end{equation}

We also calculate the expression for the second part of the perturbed Cotton tensor,
\begin{equation}
    \delta C^{(2)}_{ij} = -\frac{1}{4}\left[v_{\sigma\tau}
    \left(\Bar{g}_{j\gamma}\epsilon^{\sigma\gamma\alpha\beta}\delta\tensor{R}{^\tau _{i\alpha\beta}}
    +\Bar{g}_{i\rho}\epsilon^{\sigma\rho\alpha\beta}\delta\tensor{R}{^\tau _{j\alpha\beta}}\right)\right].
\end{equation}
We see that there can be four sets of index choices for $\sigma$ and $\tau$ which make these terms to be
nonzero. Using all four, we get an expression for the perturbation,
\begin{eqnarray}
    \delta C^{(2)}_{ij} & = & -\frac{1}{2}\left[\bar{g}_{i\rho}\epsilon^{\eta\rho zk}
    \left(v_{\eta\eta}\delta\tensor{R}{^{\eta}_{jzk}} + v_{\eta z}\delta\tensor{R}{^{z}_{jzk}}
    + v_{z\eta}\delta\tensor{R}{^{\eta}_{j\eta k}} + v_{zz}\delta\tensor{R}{^{z}_{j\eta k}}\right)\right]
    \nonumber \\
    & & -\frac{1}{2}\left[\bar{g}_{j\gamma}\epsilon^{\eta\gamma zk}\left(v_{\eta\eta}
    \delta\tensor{R}{^{\eta}_{izk}} + v_{\eta z}\delta\tensor{R}{^{z}_{iz k}}
    + v_{z\eta}\delta\tensor{R}{^{\eta}_{i\eta k}} + v_{zz}\delta\tensor{R}{^{z}_{i\eta k}}\right)\right].
\end{eqnarray}
The point to note here is that both the perturbations in Cotton tensor vanish when we chose either of the
indices $i$ or $j$ to be the direction of propagation (the $z$-direction). So the only existing Cotton tensor
perturbations can be in the $xx$, $xy$, or $yy$ components, thus mirroring the form of the perturbation
introduced into the metric.

\subsection{Non-vanishing Cotton tensor elements}

We will conclude this appendix by calculating the three specific nonzero Cotton tensor elements.
We start with the $xx$ component,
\begin{equation}
    C^{(1)}_{xx} = -\left(v_{\eta}\nabla_{z} - v_{z}\nabla_{\eta}\right)
    \left(\bar{g}_{x\rho}\epsilon^{\eta\rho z k}\delta R_{xk}\right).
\end{equation}
Using the facts that $\bar{g}_{x\rho} = \alpha^2\delta_{x\rho}$ and $\epsilon^{\eta\rho zk} = \frac{\varepsilon^{\eta\rho zk}}{\sqrt{-g}}$, we get
\begin{eqnarray}
    C^{(1)}_{xx} & = & -\frac{1}{\sqrt{-g}}\left(v_{\eta}\nabla_{z} - v_{z}\nabla_{\eta}\right)
    \left(\alpha^2\delta_{x\rho}\varepsilon^{\eta\rho zk}\delta R_{xk}\right)\\
    & = & -\frac{1}{\sqrt{-g}}\left(v_{\eta}\nabla_{z} - v_{z}\nabla_{\eta}\right)
    \left(\alpha^2\delta_{x\rho}\varepsilon^{\eta\rho zy}\delta R_{xy}\right)\\
    & = & \frac{1}{\sqrt{-g}}\left(v_{\eta}\nabla_{z} - v_{z}\nabla_{\eta}\right)
    \left(\alpha^2\delta R_{xy}\right)\\
    & = & v_{\eta}\left[\nabla_{z}\left(\frac{1}{\alpha^2}\delta R_{xy}\right)\right]
    - v_{z}\left[\nabla_{\eta}\left(\frac{1}{\alpha^2}\delta R_{xy}\right)\right].
    \label{eq:Cxx1}
\end{eqnarray}
Now the covariant derivative can be calculated as
\begin{equation}
    \nabla_{z}\left(\frac{1}{\alpha^2}\delta R_{xy}\right) = \frac{1}{\alpha^2}
    \left(\partial_{z}\delta R_{xy} - \Gamma^{\lambda}_{zy}\delta R_{x\lambda}
    - \Gamma^{\lambda}_{zx}\delta R_{\lambda y}\right).
\end{equation}
The Christoffel symbols are first order in $\gamma_{ij}$ and hence, in the linearized theory,
their contractions with $\delta R_{ij}$ may be neglected. So, the covariant derivative simplifies to
\begin{equation}
    \nabla_{z}\left(\frac{1}{\alpha^2}\delta R_{xy}\right)
    = \frac{1}{\alpha^2}\left(\partial_{z}\delta R_{xy}\right).
\end{equation}
Similarly, for the other term in eq.~\eqref{eq:Cxx1}, we have
\begin{equation}
    \nabla_{\eta}\left(\delta R_{xy}\right) = \partial_{\eta}\delta R_{xy}.
    - \Gamma^{\lambda}_{\eta y}\delta R_{x\lambda} - \Gamma^{\lambda}_{\eta x}\delta R_{\lambda y}
\end{equation}
Using the form of the Christoffel symbol $\Gamma^{i}_{\eta j}$, we can see the background contribution
(which is nonzero in this term),
\begin{equation}
    \nabla_{\eta}\left(\delta R_{xy}\right) = \partial_{\eta}\delta R_{xy} + \frac{2}{\eta}\delta R_{xy}.
\end{equation}
From this, we find
\begin{eqnarray}
    \nabla_{\eta}\left(\frac{1}{\alpha^2}\delta R_{xy}\right) & = & \frac{1}{\alpha^2}\nabla_{\eta}\delta R_{xy}
    +\partial_{\eta}\left(\frac{1}{\alpha^2}\right)\delta R_{xy} \\
    & = & \frac{1}{\alpha^2}\partial_{\eta}\delta R_{xy} + \left(\frac{2}{\eta\alpha^2}\right)\delta R_{xy}
    - \frac{2\alpha'}{\alpha^3}\delta R_{xy}\\
    & = & \frac{1}{\alpha^2}\left(\partial_{\eta}\delta R_{xy} + \frac{4}{\eta}\delta R_{xy}\right).
\end{eqnarray}
So the first part of the Cotton tensor becomes
\begin{equation}
    C^{(1)}_{xx} = \frac{1}{\alpha^2}\left(\frac{v_{\eta}}{2}\partial_{z}\delta R_{xy}
    - \frac{v_{z}}{2}\partial_{\eta}\delta R_{xy} - \frac{4 v_{z}}{\eta}\delta R_{xy}\right).
\end{equation}

In a similar fashion, for the second part we get
\begin{equation}
    C^{(2)}_{xx} = \frac{1}{\sqrt{-g}}\left[\alpha^2\delta_{x\rho}\varepsilon^{\eta\rho zk}
    \left(v_{\eta\eta}\tensor{R}{^{\eta}_{xzk}} + v_{\eta z}\tensor{R}{^{z}_{xzk}}
    + v_{z\eta}\tensor{R}{^{\eta}_{x\eta k}} + v_{zz}\tensor{R}{^{z}_{x\eta k}}\right)\right].
\end{equation}
Using the Riemann tensor components and the fact that $k$ can only be $y$, we find
\begin{equation}
    C^{(2)}_{xx} = \frac{1}{\alpha^2}\left[v_{\eta\eta}
    \left(\frac{1}{2}\partial_{z}\partial_{\eta}\gamma_{xy}\right)
    + v_{\eta z}\left(\frac{1}{2}\Box\gamma_{xy}\right) + v_{zz}
    \left(-\frac{1}{2}\partial_{\eta}\partial^{z}\gamma_{xy}\right)\right].
\end{equation}
So, the $xx$ component of Cotton tensor takes the final form
\begin{eqnarray}
    C_{xx} & = & \frac{1}{\alpha^2} \Biggl[ 
        \frac{v_{\eta}}{2}\partial_{z}\Box\gamma_{xy} 
        - \frac{v_{z}}{2}\partial_{\eta}\Box\gamma_{xy} 
        - \frac{4v_{z}}{\eta}\Box\gamma_{xy} \nonumber\\
    & & + v_{\eta\eta}\left(\frac{1}{2}\partial_{z}\partial_{\eta}\gamma_{xy}\right) 
        + v_{\eta z}\left(\frac{1}{2}\Box\gamma_{xy}\right) 
        + v_{zz}\left(-\frac{1}{2}\partial_{\eta}\partial^{z}\gamma_{xy}\right) 
    \Biggr],
\end{eqnarray}
where we have used $\Box = (\partial^2_{\eta} + 2\frac{\alpha'}{\alpha}\partial_{\eta} - \partial^2_{z})$.

To get the $yy$ component, we only need to note the facts that $\gamma_{yy} = -\gamma_{xx}$ and
$\gamma_{xy} = \gamma_{yx}$. From these follow the relation
\begin{equation}
    C_{yy} = -C_{xx}.
\end{equation}
Finally, for the $xy$ cross term in Cotton tensor we get
\begin{eqnarray}
    C^{(1)}_{xy} & = & \frac{1}{\alpha^2}\left[\frac{v_{\eta}}{2}\partial_{z}\Box\gamma_{xx}
    - \frac{v_{z}}{2}\partial_{\eta}\Box\gamma_{xx} - \frac{4v_{z}}{\eta}\Box\gamma_{xx}\right] \\
    C^{(2)}_{xy} & = & \frac{1}{\alpha^2}\left[v_{\eta\eta}
    \left(\frac{1}{2}\partial_{z}\partial_{\eta}\gamma_{xx}\right)
    + v_{\eta z}\left(\frac{1}{2}\Box\gamma_{xx}\right)
    + v_{zz}\left(-\frac{1}{2}\partial_{\eta}\partial^{z}\gamma_{xx}\right)\right].
\end{eqnarray}


\begin{thebibliography}{99}
\bibitem{ref-desitter}W. de Sitter, Proc. R. Neth. Acad. Arts Sci. \textbf{19}, 1217 (1917).
\bibitem{ref-hawking}S. W. Hawking, Commun. Math. Phys. \textbf{43}, 199 (1975).
\bibitem{ref-mukhanov}V. Mukhanov, \textit{Physical Foundations of Cosmology}
(Cambridge University Press, Cambridge, 2005).
\bibitem{ref-guth}A. H. Guth, Phys. Rev. D \textbf{23}, 347 (1981).
\bibitem{ref-linde}A. D. Linde, Phys. Lett. B \textbf{108}, 389 (1982).
\bibitem{ref-albrecht}A. Albrecht and P. J. Steinhardt, Phys. Rev. Lett. \textbf{48}, 1220 (1982).
\bibitem{ref-mukhanov_a}V. Mukhanov and G. Chibisov, JETP Lett. \textbf{33}, 532 (1981).
\bibitem{ref-starobinsky}A. A. Starobinsky, Phys. Lett. B \textbf{117}, 175 (1982).
\bibitem{ref-bardeen}J. M. Bardeen, P. J. Steinhardt, and M. S. Turner, Phys. Rev. D \textbf{28}, 679 (1983).
\bibitem{ref-planck}Y. Akrami \textit{et al.} (Planck Collaboration), Astron. Astrophys. \textbf{641}, A10 (2020).
\bibitem{ref-bicep}H. Boenish \textit{et al.} (BICEP/Keck Collaboration), Phys. Rev. Lett. \textbf{127}, 151301 (2021).
\bibitem{ref-reiss}A. G. Riess \textit{et al.}, Astron. J. \textbf{116}, 1009 (1998).
\bibitem{ref-perlmutter}S. Perlmutter \textit{et al.}, Astrophys. J. \textbf{517}, 565 (1999).
\bibitem{ref-chern}S.-S. Chern and J. Simons, Ann. Math. \textbf{99}, 48 (1974).
\bibitem{ref-jackiw}R. Jackiw and S.-Y. Pi, Phys. Rev. D \textbf{68}, 104012 (2003).
\bibitem{ref-yunes2009}N. Yunes and F. Pretorius, Phys. Rev. D \textbf{79}, 084043 (2009).
\bibitem{ref-yunes2010}N. Yunes and S. A. Hughes, Phys. Rev. D \textbf{82}, 082002 (2010).
\bibitem{ref-alexander}S. Alexander, L. S. Finn, and N. Yunes, Phys. Rev. D \textbf{78}, 066005 (2008).
\bibitem{ref-lue}A. Lue, L. Wang, and M. Kamionkowski, Phys. Rev. Lett. \textbf{83}, 1506 (1999).
\bibitem{ref-contaldi}C. R. Contaldi, J. Magueijo, and L. Smolin, Phys. Rev. Lett. \textbf{101}, 141101 (2008).
\bibitem{ref-dyda}S. Dyda, \'{E}. \'{E}. Flanagan, and M. Kamionkowski, Phys. Rev. D \textbf{86}, 124031 (2012).
\bibitem{ref-bunchdavies}T. S. Bunch and P. C. W. Davies, Proc. R. Soc. Lond. A \textbf{360}, 117 (1978).
\bibitem{ref-isaacson1}R. A. Isaacson, Phys. Rev. \textbf{166}, 1263 (1968).
\bibitem{ref-isaacson2}R. A. Isaacson, Phys. Rev. \textbf{166}, 1272 (1968).
\bibitem{ref-wen} W.~Zhao, T.~Zhu, J.~Qiao, and A.~Wang, Phys.\ Rev.\ D \textbf{101}, 024002 (2020).
\bibitem{ref-hwang1998}J. Hwang, Class. Quantum Grav. \textbf{15}, 1401 (1998).
\bibitem{ref-choi2000}K. Choi, J. Hwang, and K. Hwang, Phys. Rev. D \textbf{61}, 084026 (2000).
\bibitem{ref-turner} M. S. Turner, Phys. Rev. D \textbf{28}, 1243 (1983).
\bibitem{ref-sikivie} P. Sikivie, Lect. Notes Phys. \textbf{741}, 19 (2008)..
\bibitem{ref-marsh} D. J. E. Marsh, Phys. Rep. \textbf{643}, 1 (2016).
\bibitem{ref-alexander2009} S. Alexander and N. Yunes, Phys. Rep. \textbf{480}, 1 (2009).
\bibitem{ref-kamionkowski} M. Kamionkowski, A. Kosowsky, and A. Stebbins, Phys. Rev. D \textbf{55}, 7368 (1997).

\end{thebibliography}
\end{document}